\documentclass{article}
\usepackage{courier}
\usepackage{pstricks}
\usepackage{pst-all}
\usepackage{pst-plot}

\begin{document}

\title{RECOMBINATION AND BITSETS}

\author{JOSE RODRIGUEZ, F. B. CHRISTIANSEN\footnote{Bioinformatics Research Center,
University of Aarhus, Denmark.} \& H. F. HOENIGSBERG\footnote{ Instituto de Gen\'etica Evolutiva y Biolog\'\i a Molecular \& Instituto de Gen\'etica Ecol\'ogica y Biodiversidad del Tr\'opico Americano. Bogot\'a D.C., Colombia.}}

\date{}

\maketitle

A bitset is a set that encodes for a binary number. Bitsets are at the basis of a beautiful theory of recombination with n-loci and here we begin from scratch and advance to include     the derivation of  the fundamental results about the evolution of gamete frequencies and of disequilibrium measures with and without migration.  All techniques have been illustrated and we have invested moreover a great effort to make the mathematics of this work accessible even for students in their first year at the university.

\section{RECOMBINANT OPERATORS}

Amphimictic diploid organisms are those that develop from a single cell which results from the fusion of paternal haploid spermatozoa and maternal
haploid ovule. Each cell in the resulting organism will have approximately
the same genetic material, although as cellular subdivisions run,
especially those at gametic meiosis, linkage can change and the
individual will produce gametes with mixed maternal and paternal
genomes. The process leading to this ordered mixing is called recombination.

Diploidiness offers at least two possible evolutionary phenomena:
1) It allows for two versions of given genetic information, each one
of them can be fitted to different environments. In this case, recombination
emerges as a process that puts together harmonious gene
complexes in different arrangements. 2) Given that 
reproduction occasionally involves useless errors in replication and that these
errors are few, recombination can be viewed as a mechanism that
either repairs by reestablishing adaptive combinations or that
puts together potentially harmonious genes lying in different gametes.
In fact, this is Muller's theory of a "higher evolutionary rate" for
sexual (amphimyxis) over asexual (parthenogenesis) species. However,
Muller's theory does not consider small population sizes. The advantage
of sexual diploidy, because of its recombinational repairing of
DNA machinery, disappears in small populations because new
mutants will, through statistics, either become fixed or disappear
before they can recombine profitably.

Let us consider loci from 1 to n, each one with two alleles or versions noted
0 and 1. Then, a gamete is represented by a binary number, say, for
n=4, the binary numbers 1010, 0001 and 1101 represent three different
gametes. Recombination can be modeled by a relation $F$ so
that each ordered pair of gametes (the first place for the maternal
gamete and the second for the paternal one that fused into the
original zygote) implies a third gamete constructed with the alleles
of the ordered pair. So, $F (1010,0001)$ can be 1010 or 0001 or 1000
or 1001 or 0010 but cannot be 1100. 

Since $F$ is not univalued, a set of functions can be defined by introducing the concept of recombinant operator $F_G$, where $G$ is a   binary number, and that operates over a zygote $(G_1=maternal ,G_2=paternal)$, such that $F_G (G_1, G_2)$ is a third gamete that has at a given locus  the maternal allele whenever $G$ has a 1 or the paternal one if $G$ has a 0. For example, $F_{1011} (1001, 0011) =
1001$; $F_{1001}( 1001,0011) = 1011$; $F_{0011}(1 001, 1100) = 1101$ (see
figure 1 ).

\stepcounter{figure}
\psset{unit=0.05cm}
\begin{center}
\begin{pspicture}(0,0)(129,75)
\rput{0}(15,20){\psellipse[](0,0)(15,-10)}

\pscustom[]{\psbezier(45,65)(59,72)(68,73.5)(75,70)
(82,66.5)(83.5,61.25)(80,52.5)
(76.5,43.75)(76.5,36.25)(80,27.5)
(83.5,18.75)(79,15)(65,15)
(51,15)(46.5,18.75)(50,27.5)
(53.5,36.25)(52,47.5)(45,65)
}
\psline(1,16)(15,30)
\psline(5,20)(19,10)
\psline(11,16)(27,26)
\psline(80,40)(95,40)
\psline(90,45)(95,40)
\psline(90,35)(95,40)
\psline(30,40)(45,40)
\psline(40,45)(45,40)
\psline(40,35)(45,40)

\rput{0}(15,60){\psellipse[](0,0)(15,-10)}

\psline(1,56)(15,70)
\psline(5,60)(19,50)
\psline(12,55)(27,66)

\rput{0}(114,58){\psellipse[](0,0)(15,-10)}

\psline(100,54)(114,68)
\psline(104,58)(118,48)
\psline(111,53)(126,64)

\rput{0}(113,19){\psellipse[](0,0)(15,-10)}

\psline(99,15)(113,29)
\psline(103,19)(117,9)
\psline(110,14)(125,25)
\psline(50,51)(68,72)
\psline(55,56)(73,16)
\psline(63,37)(81,56)

\pscustom[]{\psline(2,60)(2,62)
\psline(2,62)(4,62)
\psline(4,62)(4,64)
\psline(4,64)(7,64)
\psline(7,64)(7,66)
\psline(7,66)(9,66)

}

\pscustom[]{\psline(100,19)(100,21)
\psline(100,21)(102,21)
\psline(102,21)(102,23)
\psline(102,23)(105,23)
\psline(105,23)(105,25)
\psline(105,25)(107,25)

}

\pscustom[]{\psline(8,56)(6,54)
\psline(6,54)(10,54)
\psline(10,54)(8,56)

}

\pscustom[]{\psline(106,54)(104,52)
\psline(104,52)(108,52)
\psline(108,52)(106,54)

}

\pscustom[]{\psline(56,46)(54,44)
\psline(54,44)(58,44)
\psline(58,44)(56,46)

}

\pscustom[]{\psline(49,61)(49,63)
\psline(49,63)(51,63)
\psline(51,63)(51,65)
\psline(51,65)(54,65)
\psline(54,65)(54,67)
\psline(54,67)(56,67)
}

\pscustom[]{\psline(1,21)(3,21)
\psline(3,21)(3,22)
\psline(3,22)(5,23)
\psline(5,23)(6,25)
\psline(6,25)(9,26)

}

\pscustom[]{\psline(101,60)(103,60)
\psline(103,60)(103,61)
\psline(103,61)(105,62)
\psline(105,62)(106,64)
\psline(106,64)(109,65)

}

\pscustom[]{\psline(51,59)(53,59)
\psline(53,59)(53,60)
\psline(53,60)(55,61)
\psline(55,61)(56,63)
\psline(56,63)(59,64)

}

\pscustom[]{\psline(6,17)(4,15)
\psline(4,15)(14,11)

\psline(14,11)(6,17)
\closepath}

\pscustom[]{\psline(104,16)(102,14)
\psline(102,14)(112,10)

\psline(112,10)(104,16)
\closepath}

\pscustom[]{\psline(56,32)(54,30)
\psline(54,30)(64,26)

\psline(64,26)(56,32)
\closepath}

\psline(12,60)
(10,60)
(10,62)
(16,62)
(16,64)
(14,64)(13,64)

\psline(111,19)
(109,19)
(109,21)
(115,21)
(115,23)
(113,23)(112,23)

\psline(67,62)
(65,62)
(65,64)
(71,64)
(71,66)
(69,66)(68,66)

\pscustom[]{\psline(13,20)(10,21)
\psline(10,21)(19,23)
\psline(19,23)(16,26)
}

\pscustom[]{\psline(112,58)(109,59)
\psline(109,59)(118,61)
\psline(118,61)(115,64)
}

\pscustom[]{\psline(67,55)(64,56)
\psline(64,56)(73,58)
\psline(73,58)(70,61)
}

\rput{0}(21,56){\psellipse[](0,0)(2.5,-1)}
\rput{0}(118.5,15){\psellipse[](0,0)(2.5,-1)}
\rput{0}(71,38){\psellipse[](0,0)(2.5,-1)}
\pspolygon(18,17)(23,17)(23,15)(18,15)
\pspolygon(70,33)(75,33)(75,31)(70,31)
\pspolygon(118,54)(123,54)(123,52)(118,52)
\end{pspicture}

\bigskip
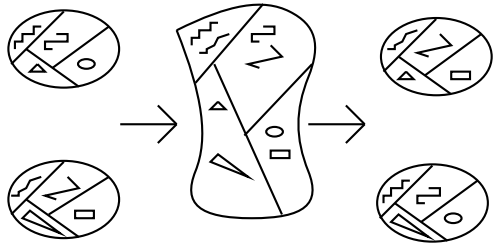
\textit{Figure \thefigure. Schematic representation of recombination: two initial gametes interact to form an exited supergamete that can split into resultant gametes. Output and initial gametes are different forms of information packing. }

\end{center}

Formally, let   $G_1$   and $G_2$ be two gametes and $G$ a binary number  that represents a recombinant operator, i.e., $G =\sum a_i 2^{i -1}$,
$G_1 =\sum b_i 2^{i -1}$, $G_2 =\sum c_i 2^{i -1}$ where sums extend from 1 to $n$. Then
$F_G (G_1,G_2)=\sum d_i 2^{i-1}$ where $d_i=b_i$ if $a_i= 1$ or   $d_i =c_i$ if $a_i=0$.
Coefficients $a_i,b_i,c_i,d_i$  are 0 or 1 for each $i$ from 1 to $n$.
That gametes and  recombinant operators are in a one to one
correspondence can be realized from the following identity:

\begin{equation}
F_G (ONE, ZERO)=G 	
\end{equation}

Where $ONE$ stands for the gamete that has allele number 1 at all loci,
and similarly, $ZERO$ stands for the gamete which has allele 0 at all
loci. For instance, $F_{1101} (1111,0000)= 1101$. By this reason, we are allowed to confound a recombinant operator with a gamete. 

To construct mathematical models it is necessary to assign to
each   recombinant operator $F_G$, which is an abstract concept
meaning recombinant power, a probabi1ity $R(F_G)$. Thanks to
equation (1) it is possible to identify this abstract concept with
measurable items. Then, it is possible to
use the frequencies $R(G)$ of each output gamete $G$ as estimators of
the probabi1ities $R(F_G)$, of the corresponding   recombinant operator
 $F_G$. The distribution of $R(G)$ on gametes, rather than on
recombinant operators, will be referred to as Geirenger recombination
distribution (Christiansen, 1987).

Physically, recombination is understood as an exchange of information
between a pair of gametes, therefore, it would be more
natural to represent the result not just as one gamete, but rather as
two, say $F_G (G_1,G_2)= (G_3,G_4)$ with the property that $G_1+G_2 =
G_3 + G_4$ where the binary sum is taken locus by locus. For example,
1001 + 1100 = 1101 + 1000, meaning that if gametes 1001 and
1100 combine and as a result of recombination gamete 1101 is
produced then it is necessary that gamete 1000 be produced also.

Note, that in general, $F_G (G_1,G_2) = G_3$ does not imply that $F_{1-G}
(G_1,G_2) = 1-G_3$. For example, $F_{1101} (0010,1100) = 0000$ while $F_{0010}
(0010,1100) = 1110$, but $0000 + 1110 \ne  ONE$. However, when
$G_1 $= ONE and $G_2 $= ZERO, then we have that if $F_G (G_1, G_2) = G_3$,
then $F_{1-G} (G_1 ,G_2) = 1-G_3$. It would be possible to model recombination
by pairs of recombinant operators $(F_{G_1}, F_{1-G_1} )$, and in this
case, the null hypothesis per excellence would be Mendel's rule of
segregation expressed by equation

\begin{equation}
 R(F_G) =R(F_{1-G}) 	
\end{equation}

or in Geirenger formulation as

 \begin{center}
\hspace{4.9cm}$R(G) = R(1 -G)$ \hfill {(2')}
\end{center}

Mendel's rule (2, 2') assumes that recombination does not interfere with
the performance of a given gamete relative to zygote formation.

\section{SOME EXAMPLES ON GEIRENGER DISTRIBUTIONS}

To fix ideas, let us consider some simple and important cases.

\subsection{ The two loci case}

Recombination is evidenced by output gametes 10 and 01, while
no recombination is demonstrated by output gametes 00 and 11.
When Mendel's law of segregation (2') is obeyed, a single parameter
$c = R (01) + R (10)$ (equals the probability of recombination)
defines the whole process. In effect, by (2'), $R(10) = R(01) = c/2$,
and given that the sum of all probabilities renders one, then $R(00)+R( 11 )=
1-c$, the probability of no recombination. Again by applying (2'), we
have $R(00)=R(11)=(1-c)/2$. Without Mendel's law of segregation
three parameters would be necessary to define recombination. The only
granted information we have is the normalizing condition $\sum R(G)= 1$.
Since recombination is not a deterministic phenomenon it must
be modeled by a random variable, with a given expected value and a
given expected variance. In the two loci case, the probability of
recombination, $c$, has a maximum expected value of 1/2, as we show
below. To recombine, gametes must interact, but since they are
physically stable DNA structures, they must be activated with some
specific energy. So, let us consider the following chemical reaction
model, as in figure 1: 

$$G_1 + G_2     \rightleftharpoons  (G_1 G_2)^* \rightarrow  G_3 + G_4$$

where $G_1, G_2$ are the gametes that gave origin to the output gametes
$G_3, G_4$, while ($G_1G_2)^*$ is the activated complex. $G_1, G_2$ have the probability $v$ to go into the activated complex $(G_1 G_2)^*$ which can
then split into output gametes $G_3, G_4$ with probability $s$. Former
gametes $G_1, G_2$ can be recovered with probability $t$. The maximum
value of $v$ is 1, while $t$ and $s$ can be considered equal to one another, since $t+s=1$,  $s=1/2$. Since the probability of recombination $c=vs$, then $c=1/2$ is at maximum. This bound could serve to evidence selection but since this 
value   has sampling variance,    calculations involving the expected variance of
$c$ must be carried out (See Karlin et altri, 1978).

\subsection{The three loci case}

In the three loci case there are eight types of recombinant gametes
$(2^3)$ whose probabilities are to be specified. Mendel's rule
provides 4 constraints: 

$R(000) = R(111)$, 

$R(001) = R(110)$, 

$R(010) = R (101)$ 

$R(011) = R(100)$. 

Moreover since we have the
condition $\sum R(G) = 1$ three more independent equations are required.
This can be reduced to $R (G)$ for three gametes lying in
different equations among the four just enumerated. But there are
other forms to produce three independent equations. We are going to
present two of them: the exclusive and the inclusive representations.

\

The exclusive formulation involves the following parameters: 
\begin{enumerate}
	\item $r$: the probability of recombination between the first two loci without
recombination between the last two $r = R (011) + R(100)$ (from left
to right). 
\item $s$: the probability of recombination between the last two
loci without recombination between the first two, $s=R (110) + R
(001)$. 
\item $t$: the probability of simultaneous recombination between
the first two loci and the last two: $t= R (101) + R (010)$. 
\end{enumerate}

The inclusive formulation and its relation to the exclusive one is
established by the following parameters:

\begin{enumerate}
	\item  $u$: the probability of
recombination between the first two loci with or without recombination
between the last two, $u=R (010) + R(101) + R (100)
+ R(011) = r + t$.
\item $v$: the probability of recombination between
the last two loci with or without recombination between the first
two, $v=R (101) + R(010) + R(110) + R(001) = t + s$.
\item   $w$: the
probability of recombination between the first two loci without
recombination between the last two, plus the probability of recombination
between the last two without recombination between the first
two, $w=R(011) + R (100) + R(110) + R(001) = r + s$.
\end{enumerate}

\subsection{The four loci case}

In the four loci case we have $2^4 = 16$ types of gametes. Mendel's
law furnishes 8 equations, with the normalizing condition $\sum R (G)= 1$
we have a total of 9 equations. Therefore, 7 other equations are
required to completely specify the ensemble of probabilities of each
kind of output gamete.

The parameters corresponding to the exclusive representation of
the recombination process can be defined in the fo1lowing way:

$E_1 = R(1000) + R(0111)$

$E_2 = R(1100) + R(0011)$

$E_3 = R(1110) + R(0001)$

$E_{12} = R(1011) + R(0100)$

$E_{13} = R(1001) + R(0110)$

$E_{23} = R(1101) + R(0010)$

$E_{123} = R(1010)+ R(0101)$.

Sub indexes indicate the division at which recombination takes place.
For example, $E_{12}$ means the probability of simultaneous recombination
between loci 1 and 2 and between loci 2 and 3.

One possible inclusive representation can be derived from the
exclusive one by changing in the second one, the connective "and" for
the connective "or"; for example, given that $E_1$ stands for the probability of recombination between loci 1 and 2 and no recombination
among any other loci, we define $I_1$ as $E_1 + E_{12} + E_{13} +
E_{123}$, i.e., the probability of recombination between loci 1 and 2 with
or without recombination among other loci, which is the same as
the probability of recombination involving locus 1. All cases are: 

$I_1 = E_{1} + E_{12} + E_{13} + E_{123}$  

$I_2 = E_{2}+ E_{12}+ E_{23}+ E_{123} $

$I_3 = E_{3}+ E_{13}+ E_{23}+ E_{123}$

$I_{12} =E_{1 }+ E_{2 }+ E_{12 }+ E_{13 }+ E_{23 }+ E_{123} $

$I_{13} = E_{1 }+ E_{3 }+ E_{12 }+ E_{13 }+ E_{23 }+ E_{123} $

$I_{23} = E_{2 }+ E_{3}+ E_{12}+ E_{13}+ E_{23}+ E_{123} $

$I_{123} =E_{1}+E_{2}+E_{3}+E_{12}+ E_{13}+E_{23}+E_{123}$

Observe that $I_{23} =I_2 + I_3$ where the sum is understood as "sum
without repetitions".

These representations can be generalized to $n$ loci, and with an
increasing $n$ it is possible to present more and more different types of
representations. However, any representation must have the same
number of independent equations: Mendel's law provides $2^n/2=2^{n-1}$
equations, which together with the normalizing condition provides a
total of $2^{n-1} + 1$ equations. Therefore, any representation must have
$2^n-2^{n-1}-1=2\times 2^{n-1}-2^{n-1}-1 =2^{n-1}-1$ independent equations.

\section{BITSETS AND RECOMBINANT OPERATORS}

Binary notation of gametes and of  recombinant operators
is particularly useful when computers are used. Gametes are represented
by vectors in a discrete space of $n$ dimensions each one with
two possible states. Nevertheless, proofs of assertions that are valid for $n$ loci, $n$ being any
number up to infinite, get considerable simplification and mathematical
beauty is improved by using set theory, for it is possible to represent
a recombinant operator or a gamete by a set that is appropriately called  bitset. Java includes inbuilt facilities to deal with bitsets (Rodriguez, 2009). A great effort has been invested to make this beautiful theory accessible to researchers and so it clarifies a previous version (Rodriguez, Christiansen and Hoenigsberg, 1988). 

Any recombinant operator   is associated with a subset of natural numbers corresponding to those loci in which the given operator picks up the maternal allele. For example, to the recombinant operator
(or   gamete)  1001 corresponds the set $\{1,4\}$ while gamete
0101 corresponds to the set $\{2,4\}$ and the set  $\{ 1,2,3 \}$ represents the recombinant mode 1110. Correspondence between binary numbers and subsets of the set $\{1, ..,n\}$ is biunivocal, for to binary numbers 1 and 0 corresponds respectively the notions "to be" and "not to be" in the representation set. Representation of gametes by bitsets was
introduced by Schnell (1961).

\

Let us review the elementary definitions and properties of set
theory. 

Always, the reference frame in set theory is the universal set,
which in our case is the set $N=\{1, ...n\}$. It is said that $A$ is a subset of $B$, noted $A < B$, if any element of A is in B too. Observe that $A < A$ for any subset of $N$.

 Operations are defined
between subsets of $N$. The union takes a pair of subsets $A$ and $B$, and
produces and output $A \cup B$, which is a subset that contains all the
elements that are in $A$ or in $B$, without repetitions. The intersection
of $A$ and $B$, $A \cap   B$, is a subset consisting of all elements common to
$A$ and $B$. The subtraction of $B$ from $A$ gives a subset $A-B$ which includes all
elements that are in $A$ but not in $B$. The symmetric difference between $A$ and $B$, gives a
subset $A \Delta B$ that equals $(A-B) \cup (B-A)$. And, finally, to any subset
$A$ is associated its set of subsets $\wp (A) =\{ B,\  B  \ is \  subset \ of \ A: A < B \}$.

Venn's representation of these operations is presented in figure 2.

When some property is to be visualized, it can be discussed on Venn's
diagrams; if 3 subsets are involved, then the discussion can be guided
by a Venn's diagram involving three circles in which there are three
two-set and one three-set intersections. Some examples follow: let
 $N =\{ 1,.. .. ,8 \}$; $A = \{ 1,2,3,4 \}$, $B = \{3,4,7 \}$ then 
 
 $A \cup B = \{ 1,2,3,4,7\}$
 
$A\cap   B= \{3,4\}$, 

$A-B=\{1,2\}$, 

$B-A= \{7\}$,

$[B-A]-B= \{ \} = \emptyset $, 
this is the empty set. 

$A\Delta B = \{1,2\} \cup \{7\}$= $\{1,2,7\}$.

$ \wp(B) = \{\emptyset$, $\{3\}$, $\{4 \}$,  $\{7\}$,
$\{3,7\}$, $\{3,4 \}$, $\{ 4,7 \}$, $\{3,4,7 \} \}$. 

Observe that $A-B$ is different from $B-A$
but $A\Delta B=B  \Delta A$.

\stepcounter{figure}
\psset{unit=0.05cm}
\begin{center}
\begin{pspicture}(0,0)(120,140)
\rput{0}(70,70){\psellipse[](0,0)(10,-10)}
\rput{0}(55,70){\psellipse[](0,0)(10,-10)}
\pspolygon[](40,90)(85,90)(85,50)(40,50)
\rput{0}(105,20){\psellipse[](0,0)(10,-10)}
\rput{0}(90,20){\psellipse[](0,0)(10,-10)}
\pspolygon[](75,40)(120,40)(120,-0)(75,-0)
\rput{0}(30,20){\psellipse[](0,0)(10,-10)}
\rput{0}(15,20){\psellipse[](0,0)(10,-10)}
\pspolygon[](0,40)(45,40)(45,0)(0,0)
\rput{0}(105,120){\psellipse[](0,0)(10,-10)}
\rput{0}(90,120){\psellipse[](0,0)(10,-10)}
\pspolygon[](75,140)(120,140)(120,100)(75,100)
\rput{0}(30,120){\psellipse[](0,0)(10,-10)}
\rput{0}(15,120){\psellipse[](0,0)(10,-10)}
\pspolygon[](0,140)(45,140)(45,100)(0,100)
\psbezier(5,120)(5,120)(5,120)(5,120)
\psline(15,130)(5,120)
\psline(6,116)(19,129)
\psline(22,127)(8,113)
\psline(11,111)(30,130)
\psline(15,110)(34,129)
\psline(23,112)(37,127)
\psline(26,111)(39,124)
\psline(30,110)(40,120)
\psline(95,120)(99,124)
\psline(96,116)(100,120)
\psline(40,85)(45,90)
\psline(40,80)(50,90)
\psline(40,75)(55,90)
\psline(40,70)(60,90)
\psline(40,65)(45,70)
\psline(55,80)(65,90)
\psline(40,60)(46,66)
\psline(59,79)(70,90)
\psline(40,55)(48,63)
\psline(62,77)(75,90)
\psline(40,50)(51,61)
\psline(64,74)(80,90)
\psline(45,50)(55,60)
\psline(50,50)(85,85)
\psline(55,50)(85,80)
\psline(60,50)(85,75)
\psline(65,50)(75,60)
\psline(70,50)(85,65)
\psline(75,50)(85,60)
\psline(80,50)(85,55)
\psline(85,70)(75,60)
\psline(6,16)(19,29)
\psline(5,20)(15,30)
\psline(8,13)(22,27)
\psline(12,11)(20,20)
\psline(15,10)(21,16)
\psline(80,20)(90,30)
\psline(81,16)(94,29)
\psline(83,13)(97,27)
\psline(86,11)(95,20)
\psline(90,10)(96,16)
\psline(99,24)(105,30)
\psline(100,20)(109,29)
\psline(98,13)(112,27)
\psline(101,11)(114,24)
\psline(105,10)(115,20)
\rput(10,130){}
\rput(10,130){}
\rput(35,135){B}
\rput(10,135){A}
\rput(110,135){B}
\rput(85,135){A}
\rput(75,85){B}
\rput(50,85){A}
\rput(35,35){B}
\rput(10,35){A}
\rput(110,35){B}
\rput(85,35){A}
\rput(25,105){$A\cup B$}
\rput(5,105){}
\rput(5,110){}
\rput(5,110){}
\rput(5,110){}
\psline(85,90)(65,70)
\rput(60,45){$A'$}
\rput(17,5){$A-B$}
\rput(95,5){$A \Delta B$}
\rput(9,106){}
\rput(98,106){$A\cap   B$}
\rput(81,107){}
\end{pspicture}

\bigskip
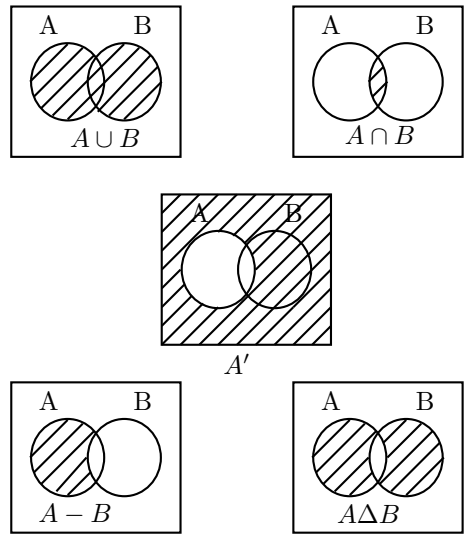
\textit{Figure \thefigure.  Venn's diagrams of set operations. Binary operation union: $\cup$; intersection:
$\cap  $; the unitary operation
complement:'; difference: -, and symmetric difference: $\Delta$  (more explanation in the text). }

\end{center}

In general, it is said that an operation $\odot$ is commutative,
if the order in which subsets are presented to the operation is not
important $A \odot B = B \odot A$. It is said that an operation  $ \odot $  is associative if   $(A \odot B) \odot C = A \odot(B\odot C)$. In this case, the parenthesis does not matter. For
example, while $(A-B)-C$ is not equal to $A-(B-C)$, $A \Delta (B\Delta C)$ equals
$(A \Delta B)\Delta C$. Union, $\cup$, intersection, $\cap  $, and symmetrical difference, $\Delta$,
are commutative and associative operations over $N$. Moreover, $A \Delta \emptyset =
(A-\emptyset) \cup (\emptyset - A) = A $ for any $A<N$, i.e., $\emptyset$ is a neutral element for $\Delta$,
an element that plays with respect to $\Delta$ the same role as 0 with
respect to the sum of integer numbers; on the other hand, $ A\Delta A=
(A-A) \cup(A-A)=\emptyset$, i.e., $A$ cancels itself with respect to the neutral
element $\emptyset$ of $\Delta$.

 Also, we have the distributive laws: 

$A\cup(B\cap   C)=
(A\cup B)\cap   (A \cup C)$ 

$A \cap   (B\cup C)=(A\cap   B) \cup (A \cap   C)$ 

which can be visualized
through Venn's diagrams or, verbalized, say, for the first law, in
the following way: elements common to $B$ and $C$ gathered with those
in $A$ are those same elements that are both in the union of $A$ and $B$
and in the union of $A$ and $C$.

The set $N-A$ is noted $A'$ and it is referred to as the complement of
A. Mendel's law is written as

\begin{equation}
 R(F_A) =R(F_{A'}) 	
\end{equation}

or in Geirenger representation as

 \begin{center}
\hspace{4.9cm}$R(A) = R(A')$ \hfill {(3')}
\end{center}

Formula (3') is sometimes the most suitable,
because it directly represents a recombination process, but when
arithmetic is involved, formula (3) is better.
Since a gamete is represented by a subset $A$ of $N= \{ 1,.. .., n  \}$
then a genotype is a couple of gametes, and given that we gave the
female gamete in the first place, a genotype is and ordered pair of
subsets $(A_1, A_2) $ belonging to $ \wp(N) X \wp(N)$ which is the set of all couples that can be formed with elements of $\wp(N)$.

\section{EVOLUTION OF GAMETE FREQUENCIES IN
A PANMICTIC POPULATION}

A panmictic population is one that has no reproductive barriers or biases with respect to random mating. 
As an introduction let us show that in a panmictic infinite population,
random mating of diploid zygotes is equivalent to the random
mating of the haploid gametes they produce. In fact, let us consider
that in the population the frequency of females with genotype $G_iG_j$
is equal to the frequency of males with the same genotype, equal to
$p_{ij}$ and that each female contributes $F$ gametes while each male
contributes $M$ gametes. Let us imagine that each individual throws its gametes into a common reservoir. Let $p_i$ be the frequency of gamete
$G_i$ and   $p_k$ the frequency of gamete $G_k$. Since each
male contributes a number of $M$ gametes, and females $F$ gametes,
then we have: 

$p_i =Mp_{ii}/ (M+F) + Fp_{ii}/ (M+F) $

$ \hspace{0.5cm}+ M\sum p_{ij}/(M+F)
+   F\sum p_{ij}/(M+F) $

$ \hspace{0.5cm} =p_{ii}+ \sum p_{ij} $

similarly

$p_k = p_{kk} +  \sum p_{kl}$

Therefore, the frequency of the haploid mating $G_iG_j$ at the gamete
reservoir is 

$p_ip_k = (p_{ii} +   \sum  p_{ij}) (p_{k k} +   \sum  p_{k l} )$

$ \hspace{0.7cm}  = p_{ii} p_{kk} + \sum  p_{ii}p_{kl} + \sum  p_{ij}p_{kk} + \sum \sum     p_{ij}p_{kl}$.

 We include in this last equation all possible matings among diploid
zygotes that will produce the zygote $G_iG_k$: matings among homozygotes,
homozygotes with heterozygotes, and among heterozygotes. In short, we have proved that:

$p_ip_k = p_{ik}$

or that random mating of diploid zygotes is equivalent to the random
mating of the haploid gametes they produce.

Nevertheless, in this proof, there is one hidden assumption: there is
no recombination anywhere! To remove this assumption we can
proceed as follows: the frequency in the population of the individual
$G_iG_j$ is $p_ip_j$, this individual will produce gamete $G_k$ with a given probability $R_{ijk}$ that depends on the recombination scheme. 

Therefore, at the reservoir, genotype $G_iG_j$ will contribute gamete $G_k$ in a
frequency $R_{ijk}p_ip_j$. But $G_k$ can come from many more genotypes,
and the frequency of gamete $G_k$ in the reservoir will be of the
form   $ \sum R_{ijk}p_{ij}$, where the sum is extended over those individuals
$G_iG_j$ that, through recombination, can produce gamete $G_k$. Then, in
the reservoir, the probability of mating among haploid gametes $G_k$ 
and $G_l$ is of the form 

$p_kp_l = (\sum R_{ijk}p_ip_j) ( \sum R_{mnl}p_mp_n) $
$
\sum R_{ijk}R_{mnl}p_ip_jp_mp_n$

 where sums extends over $ i,j,m$  and $n$. But
this last expression is equal to the probability of the double event of
mating zygotes $G_iG_j$ with those of genotype $G_m G_n$, with the condition
that they will produce through recombination gamete $G_k$ and gamete $G_l$. So

$p_kp_l  = \sum R_{ijk}R_{mnl}p_{ij}p_{mn}$

The exact form of $R_{ijk}$ could be calculated from equations given
below, although we will not write it explicitly. As before, one could
assume that males and females produce different number of gametes.

Hence, random mating among individuals is equivalent to random
mating among their gametes! Thanks to this result we will deal with
populations of gametes rather than with populations of diploid
individuals.

\

A genotype can be represented in Venn's diagrams as in figure 3.

\

Recalling that numbers appearing in set representation of a gamete
are those corresponding to maternal alleles, while paternal alleles are
those in the complement of the set, then a genotype $(A,B)$ is homozygous
for maternal alleles at the intersection $A\cap   B$, homozygous for
paternal alleles outside the set representators, i.e., at $(A\cup B)' =
A' \cap   B'$ , and it is heterozygous at $A \Delta B = (A- B) \cup  (B-A)$ in $A-B$
site loci in which $A$ has maternal alleles while B has paternal ones 
and in $(B-A)$ site  loci in which B has maternal alleles while $A$
has paternal ones. Note that $N = (A \cup  B) \cup  (A \cup  B)' = (A \cap   B) \cup 
(A \Delta B) \cup  (A \cup  B)'$. 

\

For example, if $N = \{ 1,.. .. , 5 \}$, the genotype
($\{ 1,3,5 \}$, $\{3,4,5 \}$) is homozygous for maternal alleles at loci 3 and 5.
i.e., at $\{3,5\}$ while it is homozygous for paternal alleles at the intersection
of the complements of A and B, i.e., in $\{2,4\} \cap \{ 1,2\} = \{ 2\}$, and it
is heterozygous at loci 

( $\{ 1,3,5 \}$ -$\{3,4,5 \} ) \cup (  \{3,4,5\}$ -$\{ 1,3,5 \}$) =
$\{1,4\}$. 

Here, we have that N = $\{1,...,5 \}$= $\{3,5 \} \cup \{2\} \cup \{1,4\}$ and that these three sets are pairwise disjoint, which means that they, taken by groups of two, do not have any member in common. 

\

Which conditions must there be for genotype $(B,C)$ to give a gamete $A$ as product of recombination? In more technical words, which conditions must there be for a genotype $(B,C)$ to guarantee at least one recombinant operator $F_G$, such that $F_G (B,C) = A$?

\stepcounter{figure}
\psset{unit=0.05cm}
\begin{center}

\begin{pspicture}(0,0)(150,135)
\rput(25,5){$G_1 \cap G_2$}
\rput(80,5){$(G_1 \cup G_2)'$}
\rput(130,5){$G_1 \Delta G_2$}
\rput{0}(20,25){\psellipse[](0,0)(10,-10)}
\rput{0}(20,35){\psellipse[](0,0)(0,0)}
\rput{0}(35,25){\psellipse[](0,0)(10,-10)}
\pspolygon[](5,40)(50,40)(50,10)(5,10)
\rput{0}(70,25){\psellipse[](0,0)(10,-10)}
\rput{0}(70,35){\psellipse[](0,0)(0,0)}
\rput{0}(85,25){\psellipse[](0,0)(10,-10)}
\pspolygon[](55,40)(100,40)(100,10)(55,10)
\rput{0}(120,25){\psellipse[](0,0)(10,-10)}
\rput{0}(120,35){\psellipse[](0,0)(0,0)}
\rput{0}(135,25){\psellipse[](0,0)(10,-10)}
\pspolygon[](105,40)(150,40)(150,10)(105,10)
\rput{90}(52.5,92.5){\psellipse[](0,0)(32.5,-27.5)}
\rput{90}(100,92.5){\psellipse[](0,0)(32.5,-28.75)}
\pspolygon[](5,50)(150,50)(150,135)(5,135)
\rput(50,55){$G_1$}
\rput(100,55){$G_2$}
\psline(25,25)(29.38,28.75)
\psline(25.62,21.25)(30,25)
\psline(55,35)(60,40)
\psline(55,30)(65,40)
\psline(55,25)(70,40)
\psline(55,20)(60,25)
\psline(70,35)(75,40)
\psline(55,15)(61.25,21.25)
\psline(55,10)(63.12,18.12)
\psline(74.38,33.75)(80,40)
\psline(76.88,31.88)(85,40)
\psline(85,35)(90,40)
\psline(88.75,33.75)(95,40)
\psline(92.5,31.88)(100,40)
\psline(94.38,28.75)(100,35)
\psline(95,25)(100,30)
\psline(60,10)(65.62,15.62)
\psline(65,10)(70,15)
\psline(70,10)(78.12,17.5)
\psline(75,10)(80.62,15.62)
\psline(80,10)(85,15)
\psline(85,10)(100,25)
\psline(90,10)(100,20)
\psline(95,10)(100,15)
\psline(110,25)(120,35)
\psline(110.62,20.62)(123.75,33.75)
\psline(115.62,15.62)(125,25)
\psline(113.12,18.12)(127.5,31.88)
\psline(120,15)(125.62,20.62)
\psline(129.38,28.75)(135,35)
\psline(130,25)(139.38,34.38)
\psline(128.12,17.5)(141.88,31.25)
\psline(131.25,16.25)(143.75,29.38)
\psline(135,15)(145,25)
\end{pspicture}

\bigskip
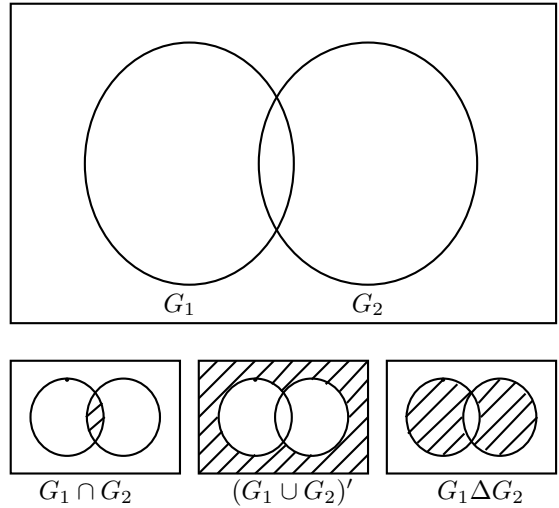
\textit{Figure \thefigure. A genotype in set representation: a genotype is a couple of gametes $(G_1, G_2)$  the first maternal and the second paternal. Since a gamete is represented by a set, the elements belonging to it are numbers corresponding to those loci in which there are maternal alleles, and elements outside the set correspond to those loci whose alleles are paternal, then a genotype is homozygous for maternal alleles in $G_1 \cap G_2$, homozygous for paternal alleles in  $(G_1 \cup G_2)'$ and heterozygous in $G_1 \Delta G_2 $. } 

\end{center}

 To fix ideas let us take $N = \{1,.. ..,6\}$ and $A = \{3, 4\}$, then the genotype $(\{2,3\}, \{4,5\})$ can recombine to give $A$, because recombinant operator  $F_{\{1,3,5\}}$ working over that genotype renders A. To see this  it is helpful to turn to binary representation: $A = \{3, 4\}$ is represented by 001100, the maternal gamete $\{2,3\}$ is denoted 011000, and the paternal gamete $\{4,5\}$ is represented by 000110, the recombinant operator  $F_{\{1,3,5\}}$  is $101010$. Therefore, we can write equation $F_G (B,C) = A$ as $F_{101010 } (011000,000110) = 001100$. Note that recombinant operators  $F_{001011 }$, $F_{101011}$ can serve too; these  modes of recombination can be easily generated in view of the fact that the genotype is homozygous for loci 1 and 6, and that therefore, at these loci, it does not matter which allele is picked up.
But, at heterozygous loci there is only one possible allele to be
picked up correctly.

Some problems on recombinant operators cannot be solved: for
example, the gamete $\{ 1, 3, 4\}$ cannot result from recombination
between gametes of genotype ($\{2,3 \}$, $\{ 4,5 \}$), i.e., there is not recombinant operator $F_G$ such that $F_G ( \{2,3 \}, \{ 4,5\}$) gives $\{ 1,3,4 \}$.  This can
be seen from binary notation: we must find $F_G$ such $F_G (011000,
000110) = 101100$, but at locus number one any recombinant operator
wou1d produce a 0 while 1 is required; in spite of this, with remaining
loci there is no problem.

In general, an individual with genotype $(B,C)$ can produce a gamete
$A$ if and only if, first, $A< B \cup  C$, for otherwise $ B$ and $C$ would not have all the maternal alleles required to construct $A$. And, second, the paternal alleles of $A$ are in $A'$, while the possible paternal alleles of $B$ and $C$ are in $B'$ or $C'$, so  we must have that 
$A' < B' \cup  C' = (B \cap    C)'$. So long as
we have that $D' < E'$ if and only if $D > E$, for any subsets $D$ and $E$,
then, condition $A'< B'\cup C' = (B \cap    C)'$ can be reexpressed as $B\cap   C < A$. 
 
 In
conclusion, genotype $(B,C)$ can recombine to give $A$ iff

\begin{equation}
	B\cap C < A < B \cup C
\end{equation}

See figure 4, in which the different areas are indicated as numbers 1-6.

 \stepcounter{figure}
\psset{unit=0.08cm}
\begin{center}

 \begin{pspicture}(0,0)(80,54.38)
\rput{0}(35,29.38){\psellipse[](0,0)(15,-15)}
\rput{0}(55,29.38){\psellipse[](0,0)(15,-15)}
\pspolygon[](10,54.38)(80,54.38)(80,4.38)(10,4.38)
\pspolygon[](30,40.62)(60,40.62)(60,18.12)(30,18.12)
\psline(30,34.38)(36.88,40.62)
\psline(30,29.38)(41.25,40.62)
\psline(30,24.38)(46.88,40.62)
\psline(30,19.38)(51.88,40.62)
\psline(34.38,18.12)(55.62,40.62)
\psline(39.38,18.12)(60,39.38)
\psline(44.38,17.5)(60,34.38)
\psline(48.75,18.12)(60,29.38)
\psline(53.12,18.12)(60,24.38)
\rput(24.38,47.5){$B$}
\rput(16.88,36.25){}
\rput(65,47.5){$C$}
\rput(13.12,29.38){1}
\rput(26.5,30){2}
\rput(34.38,30.62){3}
\rput(44.38,30.62){4}
\rput(55.62,30){5}
\rput(65,30.62){6}
\rput(43.75,11.25){$A$}
\end{pspicture}

\bigskip
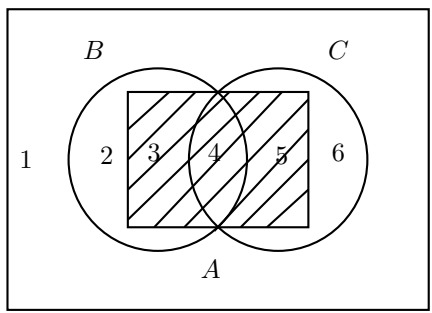
\textit{Figure \thefigure.  A genotype $(B,C)$ to produce a given gamete A as a result of recombination
must observe the following relation: $ B \cap C < A< B\cup C$. In effect, to contribute maternal alleles $A$ must be contained in $B \cup  C$. But to contribute paternal alleles, $A'$ must be contained in $B '\cup C' =(B\cap C)'$  or equivalently, $B \cap C$ must be contained in $A$. } 

\end{center}

Numbers in figure (4) have two possible interpretations: they are loci, and serve
as a particular example, or, they are ensembles of loci and serve as
general case. Having in mind that B and C are respectively female and
male gametes, these six areas are characterized by:
\begin{itemize}
	\item Area 1: the genotype is homozygous for allele 0; the recombinant
operator does not matter.
\item Area 2: the genotype $(B,C)$ is heterozygous, the mother $B$ has allele 1 and the father $C$ has allele 0, while $A$ has allele 0. The recombinant operator must, in region 2, pick allele 0
from the father, therefore the recombinant operator in region
2 is defined by a 0.
\item Area 3: the genotype is heterozygous, the mother has 1, the father 0
and the recombinant 1, so, the recombinant operator must be 1
to take the allele 1 from the mother.
\item Area 4: the genotype is homozygous with allele 1; the recombinant
operator does not matter, it can be 0 or 1.
\item Area 5: the genotype is heterozygous, the recombinant operator must
be 0 to take the allele 1 from the father.
\item Area 6: the genotype is heterozygous, the mother $B$ has a 0 whereas
the father has a 1, since the recombinant has a 0, the recombinant
operator must be 1 to pick the maternal allele.
\end{itemize}

To summarize, the recombinant operator can be noted, in binary notations,
as $F_{-01-01}$,  where the - sign stands for "either 0 or 1". On the
other hand, recombination matters only in the heterozygous areas
2, 3, 5 and 6 or in \\ $(B-C) \cup  (C-B) = B\Delta C$. In short, this can be presented in a single equation:

\begin{equation}
F_{-01-01} (011 1 00,000111) =  001110
\end{equation}

where gamete B is represented by binary number 011100 corresponding
to areas 2, 3 and 4, while gamete C is represented by 000111
corresponding to areas 4, 5 and 6. The recombinant gamete A is represented by 001110 corresponding to regions 3, 4 and 5. Moreover, the
recombinant operator  $F_{-01-01}$ picks maternal alleles at regions 3 and 6
which join to form the set $(A-C) \cup  (C-A) = A \Delta C$. To continue, we must specify how to relate equation (5) to probabilities. In other
words, knowing the probability of each recombinant operator, which is
the probability of  $F_{-01-01}$? That probability will be noted as 
$R_{\{3 ,6\} /  \{2,3,5,6\}}$ and we have:

$R_{\{3 ,6\} /  \{2,3,5,6\}}= R_{A\Delta C / B \Delta C}$

and it must be equal to $R_{001001}+ R_{001101}+ R_{101001}+ R_{101101}$. In set
notation, this can be written as

$R_{\{3,6\}/ \{2,3,5,6\}} = R_{\{3,6\}} + R_{\{3,4,6\}} + R_{\{1,3,6\}} + R_{\{1,3,4,6\}}$

Observe that any subindixal set in the left side of this equation can
be written as $\{3,6\} \cup  B$, where $B$ is a subset of $\{1,4\}  = \{2,3,5,6\}'$.
Moreover, $B$ scans   $ \wp (\{2,3,5,6\}')$, which contains all subsets of  $  \{2,3,5,6\}'$. On the other hand, in the right
side, $\{3,6 \}$ is a subset of $\{2,3,5,6 \}$. Therefore, we define,
\begin{equation}
	R(F_{B/A}) = R_{B/A} = \sum_{C<A'} R(F_{B\cup C})   \hspace{1cm} for \hspace{1cm}   B<A 
\end{equation}
 
or in Geirenger representation,

 \begin{center}
$$\hspace{3cm} R(B/A) = \sum_{C<A'} R(B\cup C) \hspace{1cm}   for \hspace{1cm} B<A  \hspace{1cm} \hfill {(6')}$$
\end{center}

$R_{B/A}= R(B/ A)$ is known as the marginal recombination distribution
of $B$ with respect to loci in $A$. In particular we have that $R_{A\Delta C / B \Delta C}$ is the marginal recombination distribution of $ A \Delta C$  with respect to $B\Delta C$, and corresponds to the condition of formation of $A$ from $(B,C)$.

With this notation we can
return to the dynamics of gamete frequencies: to find the frequency
of gamete $A$ in the offspring generation it is necessary to find the
probability of formation of a given zygote whose genotype is $(B,C)$
such that $ B \cap    C < A < B \cup  C$, and then multiply this probability by
the marginal recombination distribution of $ A \Delta C$  with respect to $B\Delta C$,
which is noted as $R_{A \Delta C/B\Delta C }$. In a panmictic population, the probability
of formation of zygote with genotype $(B,C)$ is, by definition, the
product of the frequencies in the population of $B$ and $C$, $p(B)$, $p(C)$
respectively. Therefore, the frequency of gamete $A$ in the offspring
generation in a panmictic population, $p'(A)$, is

\begin{equation}
p'(A) = \sum_{B\cap C < A < B\cup C} R_{A\Delta C/B \Delta C}  p(B) p(C) 	
\end{equation}

Equation (7) is general and valid for any number of loci.
\

Likewise to the concept of marginal recombination distribution
of the recombinant operator $B$ with respect to loci in $A$, $R_{B/A}$, a
similar definition of the marginal gamete probability  of a gamete $B$
relative to loci in $A$, $p(B/ A)$, where $ B < A$, can be given by

\begin{equation}
p (B/A) = \sum_{C<A'} p(B\cup C)
\end{equation}
 
The more interesting property of $R_{B/A}$ is that

\begin{equation}
	\sum_{B<A} R_{B/A}  = 1
\end{equation}

an equation that can be demonstrated by the substitution of $R_{B / A}$
and a change of variable (see figure 5):

\stepcounter{figure}
\psset{unit=0.05cm}
\begin{center}

\begin{pspicture}(0,0)(100,70)
\rput{0}(30,40){\psellipse[](0,0)(10,-10)}
\rput{0}(35,35){\psellipse[](0,0)(25,-25)}
\rput{0}(80,30){\psellipse[](0,0)(10,-10)}
\pspolygon[](0,0)(100,0)(100,70)(0,70)
\rput(55,60){A}
\rput(40,50){B}
\rput(90,40){C}
\rput(110,65){N}
\end{pspicture}

\bigskip
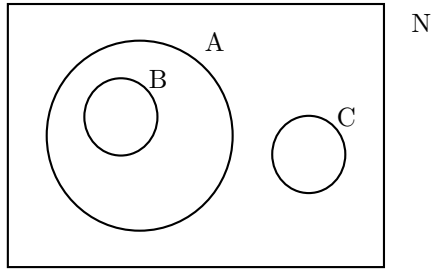
\textit{Figure \thefigure. The conditions $B<A$ and $C<A'$ are equivalent to $D<N$ where $D = B\cup C$. } 

\end{center}

$$\sum_{B<A}  R_{B/A} = \sum_{B<A} \hspace{0.2cm} \sum_{C<A'} R(B \cup C) =\sum_{D<N}  R(D) = 1$$
 
 where the pair of variables $B$ and $C$ are replaced for by the single variable $D$ which scans over all $N$, given that $B$ scans over $A$ and $C$ over $A'$ and that $A \cup A'= N$. A similar equation for $p(B/A)$ holds:

\begin{equation}
	\sum_{B<A}  p(B/A) = \sum_{B<A}  \sum_{C<A'} p(B \cup C) =\sum_{D<N}  p(D) = 1
\end{equation}
 
 These equations justify the adjective of marginal recombination
distribution and probability given to $R_{B/A}$ and to $p(B/ A)$ respectively.

\section{CARDINALITY AND PARITY OF SETS}

In order to simplify the study of the dynamics of gamete frequencies,
we need some definitions: the cardinal of a set $A$, noted as
$\#(A)$, is the number of elements in $A$. The parity of $A$, $\S (A)$ is
defined as $( -1 )^{\#A}$ , i.e., $\S  (A)$ is 1, if $\#A$ is even and -1 if $\#A$ is odd (parentheses are omitted when confusion is unexpected).

 The cardinal
and parity have the following properties:
\begin{enumerate}
	\item 
\begin{equation}
	\#(\emptyset   ) = 0, \S  (\emptyset   ) = 1
\end{equation}
 
	\item
\begin{equation}
	 \#(\wp(A)) = 2^{\#A} 
\end{equation}

	\item 
\begin{equation}
	\sum_{B<A} \S  (B) \hspace{0.2cm} equals\hspace{0.2cm}  0 \hspace{0.2cm} if  \hspace{0.2cm} A \ne \emptyset \hspace{0.2cm} and \hspace{0.2cm} 1 \hspace{0.2cm} if \hspace{0.2cm} A = \emptyset  
\end{equation}

\item 
\begin{equation}
	\sum_{C<A} \S  (C\cap B) \hspace{0.2cm} equals\hspace{0.2cm}  0 \hspace{0.2cm} if  \hspace{0.2cm} A\cap B  \ne \emptyset \hspace{0.2cm} and \hspace{0.2cm} 2^{\#A} \hspace{0.2cm} otherwise.  
\end{equation}

  \item 
\begin{equation}
	 \S  (A \Delta  B) = \S  (A \cup  B) = \S  (A) \S  (B) \hspace{0.2cm} if \hspace{0.2cm} A \cap B = \emptyset.
\end{equation}

  	\item 
\begin{equation}
	\S  (A\cup B) \S (A\cup C) = \S  (B\cup C) \hspace{0.2cm}  whenever \hspace{0.2cm}  A,B,C \hspace{0.2cm}  are \hspace{0.2cm}  disjoint
\end{equation}

	\item 
	
\begin{equation}
	\S  (A\cap B) \S (A\cap C) = \S  (A \cap(B\Delta C)).
\end{equation}

\end{enumerate}

These properties can be proved as follows: 

\begin{enumerate}
	\item The number of
elements in $\emptyset$ is zero, which is an even number, therefore, the parity
of $\emptyset$  is one.

\item  The number of subsets of a given set $A$ is $2^n$ where $n$ is
the number of elements in $A$, because, first, there are 2 possibilities
for each of $n$ places or bits in a binary number, i.e., $2^n$ binary numbers at
all, and, second, it is possible to construct a biunivocal relation
between binary numbers of $n$ bits and subsets of a set of $n$ elements.
For example, let $A = \{1,.. ..,8\}$ then to the subset $\{4,5\}$ corresponds
the binary number 00011000, and to the subset $\{2,4,6,8 \}$
corresponds the binary number 01010101 and so on.

\item To demonstrate the third property, we begin with the following
equation: $0 = (-1 + 1)^n = \sum   C(n,k) (-1)^k 1^{n-k}= \sum   C(n,k) (-1)^k$,
where the first and second sums extend from 0 to $n$, and $C(n,k)$ is
the binomial coefficient. A change of variable can be made: $(-1)^k$ is
the parity of a set $B$ of $k$ elements, and $C(n,k)$ is the number of
subsets of $k$ elements chosen among a set $A$ of $n$ elements, therefore
$\sum C (n,k) (-1)^k = \sum_{B<A}   C (n, \#   B) \S B$, where the sum in the right side
runs $k$ from 0, the cardinal of $\emptyset$, to $n$, the cardinal of $A$.

 Now,
$\sum C (n,k) \S  B$ can be written simply as $\sum    \S  E$, where the $E$ scans
over all subsets of $A$, i.e., over $\wp(A)$, and from the beginning we
knew that this sum equals zero. In this demonstration, $n$ can take
any value different from zero, i.e., $A\ne \emptyset$, for in this case $(-1 + 1)^0 =
0^0$ which is an undefined form which can be easily evaluated: if
$A= \emptyset$ , then $\sum     \S  B$, where the sum is extended over $B<A$, reduces to
$\S  (\emptyset) = 1$.

\item To demonstrate (14), we need to calculate $\sum_{C<A} \S  (C\cap  B)$. So, let us follow  figure 6:   with the definitions 

$D = C - B$,  

$E=C \cap B$,

$C$ can be partitioned in $C = D \cup E$. Therefore,  conditions $A<N$,   $C<A$, $B<N$ are equivalent to  $D<A-B$, $E<(A\cap B)$.

Hence, 
$\sum_{C<A} \S  (C\cap  B) = \sum_{D<  A -B} \sum_{E<  A  \cap B}\S  E$.

If $A\cap B = \emptyset$,  $\sum_{E<  A  \cap B}\S  E = 1$, by the previous property. In that case, $\sum_{D<  A -B} \sum_{E<  A  \cap B}\S  E = \sum_{D<  A } 1 =  2^{\#A}$ because  in $A$ there are $2^{\#A}$ subsets, then this sum equals  $2^{\#A}$; this
includes the case in which $A = \emptyset$.

But, if $A\cap B \ne \emptyset$,  $\sum_{E<  A  \cap B}\S  E = 0$, by the previous property. In that case, $\sum_{D<  A -B} \sum_{E<  A  \cap B}\S  E = \sum_{D<  A } 0 =  0$

\item Let us demonstrate now that if $A\cap B=\emptyset$  then $\S  (A\Delta B)= \S  (A\cup B)=
\S  (A) \S  (B)$. If $A$ and $B$ are disjoint then $A \Delta B = (A-B) \cup  (B-A) =
A\cup B$, and $\#   (A \cup  B) = \#    (A) + \#    (B)$. Therefore, 

$\S  (A\cup  B) =(-1)^{\#(A \cup B)}= (-1)^{(\#A +\# B)}= (-1)^{\#(A)} (-1)^{q\#B)}= \S  (A) \S  (B)$.

From this follows the sixth property:
 
\item If $A, B,C$ are pairwise disjoint,   
then $\S  (A \cup  B) \S  (A \cup  C) = \S  (A) \S  (B) \S  (A) \S  (C)$, but   this equals $\S^2 (A) \S  (B\cup C) = \S  (B \cup  C)$ for $\S  (A)$ equals
either +1 or -1. 

\stepcounter{figure}
\psset{unit=0.09cm}
\begin{center}

\begin{pspicture}(0,0)(100,80)
\rput{0}(40,40){\psellipse[](0,0)(10,-10)}
\rput{0}(35,45){\psellipse[](0,0)(25,-25)}
\rput{0}(65,45){\psellipse[](0,0)(25,-25)}
\pspolygon[](0,10)(100,10)(100,80)(0,80)
\rput(10,70){A}
\rput(90,70){B}
\rput(30,50){C}
\rput(105,75){N}
\rput(15,15){$D= C-B$}
\rput(40,15){$E = C\cap B$}
\rput(70,15){$C = D \cup E$}
\rput(10,70){}
\psline(31.25,45)(40,45)
\psline(30,40)(40.62,40)
\psline(31.88,35)(41.88,35)
\rput(40,45){}
\psline(40,45)(50,40)
\psline(40,40)(48.75,35)
\psline(41.88,35)(46.25,31.88)
\end{pspicture}

\bigskip
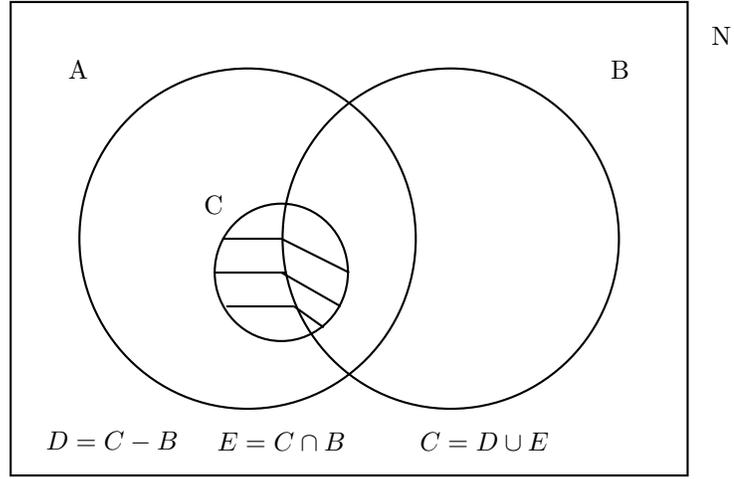
\textit{Figure \thefigure. The conditions $A<N$,   $C<A$, $B<N$ are modified by the change of variables $D = C - B$,  $E=C \cap B$, $C = D \cup E$ into  $D<A-B$, $E<(A\cap B)$. } 

\end{center}

\item To demonstrate that $\S  (A \cap B) \S  (A \cap C) = \S 
(A \cap (B\Delta C)$, which is the 7th property, let us decompose $A \cap B$ as

$A \cap B$ = $((A \cap B)-(A\cap B\cap C)) \cup  (A\cap B \cap C)=((A\cap B)-C) \cup  (A\cap B \cap C)$ 

while

$A \cap C$ =$((A \cap C)-(A\cap B\cap C)) \cup  (A\cap B \cap C)=((A\cap C)-B) \cup  (A\cap B \cap C)$.

As these unions are disjoint, then 

$\S (A \cap B)= \S (( A\cap B)-C) \cup  (A\cap B\cap C))=
\S  ( (A\cap B)-C) \S  (A\cap B \cap C)$ 

and

$\S  (A\cap C) = \S  ( (A\cap C)-B) \cup  (A\cap B\cap C))
= \S ((A \cap C)-B) \S  (A\cap   B \cap C)$.

Therefore

 $\S  (A\cap B) \S  (A\cap C) = \S ((A\cap B)-C) \S  (A\cap B \cap C) \S ( ( A\cap C)-B) \S (A\cap B \cap C) = \S ((A\cap B)-C)
\S ((A\cap C)-B) = \S ((A\cap B)-C) \cup  (A\cap C)-B) = \S  (A \cap ( B\Delta C))$,

which is an equation valid without restrictions (please, draw a Venn's diagram with three circles to make everything clear).

\end{enumerate}

\section{TRANSFORMED FREQUENCIES}

To simplify calculations on the dynamics of gamete frequencies,
the concept of transformed frequency of gamete $A$ may be worthy:

\begin{equation}
	t(A) = \sum_{B<N}    \S  (A \cap    B) p (B)
\end{equation}
 
The substitution $X=(A\cap B)$, with $X < A$, and $Y= (B-A)$, with $Y<A'$ and $B = (A\cap B) \cup (B-A) = X U Y$  gives place to a change of variable: 

$$t (A) = \sum_{X<A, Y<A' } \S  (X) p(X \cup Y) = \sum_{ X<A} \S(X) \sum_{Y<A'}  p(X \cup Y)$$

The domains of $X$ and $Y$  should be understood as: if $B$ runs over all
subsets of $N$, then $X=(A \cap    B)$ runs over all subsets of $A$, while
$Y=B-A$ can contain any element not in $A$, therefore, $Y$ moves upon
$A'$. Now, recalling definition (8):

$$p (B/A) = \sum_{C<A'} p(B\cup C)$$

we have at last that:

\begin{equation}
	t(A) = \sum_{ X<A} \S(X) \sum_{C<A'}  p(X \cup C)  =\sum_{X<A}  \S  (X)p (X/A)
\end{equation}

Here, $t(A)$ is the transformed frequency of gamete $A$ in matriarcal set
notation, $\S  (X)$ is the parity of gamete $X$, and $p (X/ A)$ is the marginal
probability of $X$ with respect to loci in $A$.

What interest would the transformed frequencies have if  it would not be possible to restore normal frequencies from
them? Therefore,
the following equalities are welcome:

\begin{equation}
	p (A)= 2^{-n} \sum    \S  (A \cap   B) t(B)
\end{equation}

\begin{equation}
	p (B/A)= 2^{-\#A} \sum_{C<A}    \S  (B \cap   C) t(C) \hspace{0.2cm} given \hspace{0.2cm} that \hspace{0.2cm} B<A
\end{equation}

Since

\begin{equation}
p(A/N)	 = \sum_{D<N'}  p(A \cup  D)  = p (A \cup  \emptyset   )= p(A) 
\end{equation}

then (20) is a special case of (21), a reverse identity that will now be
shown beginning with its right side multiplied by $2^{-\# A}$, and with a
substitution of $t(B)$ by its definition (18), 

\

$\sum_{C<A} \S(B\cap C)t(C) = \sum_{C<A} \S(B\cap C) \sum_{D<N }\S(C\cap D)p(D)$

$\hspace{3.3cm} =  \sum_{D<N} p(D) \sum_{C<A} \S(B\cap C) \S(C\cap D)$

$\hspace{3.3cm} = \sum_{D<N} p(D) \sum_{C<A} \S(C\cap (B \Delta  D) $

\
 
where we have recalling (17): 

$\S(B\cap C) \S(C\cap D) = \S(C\cap B) \S(C\cap D) = \S(C\cap (B \Delta  D) $

\

The terms of the form $\sum_{C<A}    \S  (C \cap  (B\Delta D) )$  vanish according to
(14)  if $(B\Delta D) \cap A  \ne \emptyset$.  So, we are left with those terms for which $(B\Delta D) \cap  A  = \emptyset$. Let us keep in mind that  by (21), $B<A$ and   $C<A$. 

When  $(B\Delta D) \cap  A  = \emptyset$, then $(B \Delta D) <A'$ and the most general realization of  $D$ is   $D = B \cup  L$ for $L < A'$. 
Remembering that $B$ and $L$ are disjoint because $B<A$ and   $L<A'$, we verify that 

$(B\Delta D) \cap  A  =(B\Delta  (B\cup L )) \cap  A = (B\Delta B\Delta L) \cap  A = (\emptyset \Delta L)\cap  A =L\cap  A = \emptyset$. 

Now, according to (14), $\sum   \S  (C \cap  (B\Delta D))$, sum over
$C<A$, equals $2^{\#A}$ whenever $(B\Delta D) \cap  A = \emptyset$. Using
this, replacing $D$ by $B\cup L$, and recalling definition (8) we have:

$\sum_{D<N }p(D)\sum_{C<A} \S  (C \cap (B\Delta D)) = \sum_{D<N }p(D) 2^{\#A}  $

$\hspace{3.3cm}= 2^{\#A} \sum_{ L<A'}p(B\cup L) =   2^{\#A}  p (B/A)$

which is the left side of (21) multiplied by  $2^{\#A}$  as required.

\

The most simple calculation of a transformed frequency according to (19) and (8) is that of

\begin{equation}
	t(\emptyset   ) = \sum_{X<\emptyset} \S(X)p(X/\emptyset) = \S  (\emptyset   ) p (\emptyset   /\emptyset   ) = \sum_{B<\emptyset' }p (\emptyset   \cup B) = \sum_{B<N } p (  B) = 1
\end{equation}

this means that $t (\emptyset  )$ does not evolve over time.

Transformed frequencies of one locus are very important. We
note by $t_a$ the transformed frequency $t ( $\{a\}$ )$, which calculated according
to ( 19) becomes:

\begin{equation}
	t_a=t(\{a\})=\S \emptyset   p(\emptyset / \{a\}) + \S  (\{a\}) p(\{a\} / \{a\})=p (\emptyset / \{a\})-
	p(\{a\}/\{a\}) 
\end{equation}

This $t_a$ restores marginal frequencies according to (21) as

\begin{equation}
		p (\emptyset / \{a\}) =  2^{-1} (\S \emptyset t(\emptyset)+ \S  (\emptyset)t(\{a\}))=(1/2)(1 +t_a)
\end{equation}

$\hspace{1.2cm}		p(\{a\}/\{a\})= 2^{-1}  (\S \emptyset t(\emptyset )+ \S  (\{a\})t(\{a\}))=1/2(1-t_a) \hspace{1.7cm} (25') $
	
Equations (25) and (25') will be referred to as (25).

\section{ EVOLUTION OF TRANSFORMED GAMETE FREQUENCIES}

Equation (7), 
$$
p'(A) = \sum_{B\cap C < A < B\cup C} R_{A\Delta C/B \Delta C}  p(B) p(C) 	
$$
 
 which describes the evolution of gamete frequencies,
can be transformed into a more tractable equation:

\begin{equation}
t' (A) =\sum_{B<A} R_{B/A}t (B)t(A-B)	
\end{equation}

where $t' (A)$ is the transformed frequency of the gamete $A$ in the
offspring generation, and, $t(B)$, $t(A-B)$ are respectively
the transformed frequencies of $B$ and $A-B$  in the given generation   as defined
by (18) or (19); $R_{B/A}$ is the marginal recombination distribution of
$B$ relative to loci in $A$ related to $R_A$ by (6).

\

The rest of this section will be devoted to the proof of (26),
whose demonstration is required not for any transformed frequency
$t(A)$ but for $t(N)$ only. This results from the fact that in ( 19) the
marginal gamete frequencies $p (B/ A)$ are invoked for the case in
which $B<A$, therefore $A$ plays in (26), the role of the universal set
$N$. 

\

Hence, we would like to demonstrate that 

 \begin{center}
$$\hspace{3cm} t'(N)=\sum_{B<N }R_B t(B)t(B')   \hspace{4cm} \hfill {(26')}$$
\end{center}

because $R_{B/N} = R_B$ and $t(A-B)$ is noted as $t (B')$.

\

By the definition of transformed frequencies (18) we have that

\

$t' (N) = \sum_{A <N} \S  (A \cap N) p' (A) = \sum_{A <N} \S  (A) p' (A)$ 

\

Recalling (7) and adopting
Geirenger's notation, we have:

$$
t'(N)=\sum_{A<N }\S(A) \sum_{B\cap C < A < B\cup C } R(A\Delta C/B\Delta C) p(B)p(C) 
 $$

To transform this equation into (26), terms $p (B) = p(B/N)$ and $p(C)=p(C/N)$ need to
disappear to  give their places to transformed frequencies, so they
can be replaced by their values according to inverse relations (21).

\

$
t'(N)=\sum_{A<N }\S(A) \sum_{B\cap C < A < B\cup C } R(A\Delta C/B\Delta C) \times$  

$\hspace{4cm} (2^{-n} \sum_{X< N} \S(B\cap X)t(X)) (2^{-n} \sum_{Y< N} \S(C\cap Y)t(Y))$

\

$
t'(N)=2^{-2n}\sum_{A<N } \sum_{B\cap C < A < B\cup C } \sum_{X< N} \sum_{Y< N}\S(A) R(A\Delta C/B\Delta C) \times $

 $\hspace{6.6cm}  \S(B\cap X) \S(C\cap Y)t(X) t(Y) $

Since $A$, $X$ and $Y$ are independent variables, it is possible to reorder this sum to get:

\begin{equation}
	 t' (N) = 2^{-2n} \sum_{ X<N} \sum_{Y<N }    W (X,Y) t (X) t (Y)
\end{equation}
 
where

\begin{equation}
	W (X,Y)=  \sum_{A<N } \sum_{B\cap C < A < B\cup C }  \S(A)  \S(B\cap X) \S(C\cap Y) R(A\Delta C/B\Delta C) 
\end{equation}

To evaluate $W (X, Y)$, we partition $A,B$ and $C$ into five disjoint components
$E,F,G,H,K$ such that (see figure 7): 

 \stepcounter{figure}
\psset{unit=0.08cm}
\begin{center}

 \begin{pspicture}(0,0)(80,54.38)
\rput{0}(35,29.38){\psellipse[](0,0)(15,-15)}
\rput{0}(55,29.38){\psellipse[](0,0)(15,-15)}
\pspolygon[](10,54.38)(80,54.38)(80,4.38)(10,4.38)
\pspolygon[](30,40.62)(60,40.62)(60,18.12)(30,18.12)
\psline(30,34.38)(36.88,40.62)
\psline(30,29.38)(41.25,40.62)
\psline(30,24.38)(46.88,40.62)
\psline(30,19.38)(51.88,40.62)
\psline(34.38,18.12)(55.62,40.62)
\psline(39.38,18.12)(60,39.38)
\psline(44.38,17.5)(60,34.38)
\psline(48.75,18.12)(60,29.38)
\psline(53.12,18.12)(60,24.38)
\rput(24.38,47.5){$B$}
\rput(16.88,36.25){}
\rput(65,47.5){$C$}
\rput(26.5,30){F}
\rput(34.38,30.62){G}
\rput(44.38,30.62){H}
\rput(55.62,30){E}
\rput(65,30.62){K}
\rput(43.75,47.25){$A$}

\rput(24.38,10){$T = G \cup K$}
\rput(64.62,10){$M = F \cup E$}

\end{pspicture}

\bigskip
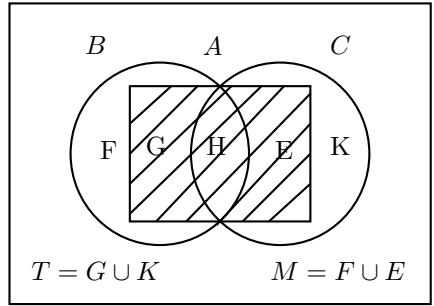
\textit{Figure \thefigure.   To calculate $W (X,Y)$ in (28), we need to partition superset $B,A,C$
into five disjoint subsets as this figure shows. } 

\end{center}

$A = E\cup G\cup H, B=F\cup  G
\cup   H, C= H\cup  E\cup  K$. 

Therefore, 

$A \Delta C = (E\cup  G\cup  H) \Delta  (H\cup  E\cup  K)$

$\hspace{1cm} = (E\cup  G\cup  H -H\cup  E\cup  K) \cup   (H\cup  E\cup  K -E\cup  G\cup  H) = G\cup  K$, 

and 

$B\Delta C =
(F\cup  G\cup  H-H\cup  E\cup  K) \cup   (H\cup  E\cup  K -F\cup  G\cup  H)$

$ \hspace{1cm} = F\cup  G\cup  E\cup  K = G \cup   K
\cup   F \cup   E$. 

If we note $T = G\cup  K = G\Delta  K$ and $M= F \cup   E = F\Delta  E$
 then

$ A \Delta  C = T$ and $B\Delta C = T \cup   M$. Applying definition (6'), we get:

\begin{equation}
	R(A \Delta C/B\Delta C) = \sum_{I <(B\Delta C)'} R (A \Delta C \cup   I) = \sum_{  I<(T\cup  M)'} R (T\cup  I )
\end{equation}

On the other hand :

$
\S (A) \S (B\cap X) \S  (C\cap Y)$

$= \S  (E\cup  G\cup  H) \S  ((F\cup  G\cup  H) \cap X) \S ( (H\cup  E\cup  K)\cap Y)$

$= \S  (E\Delta G\Delta H) \S  ((\emptyset \Delta F\Delta G\Delta H)\cap X) \S ((\emptyset \Delta H\Delta E\Delta K) \cap Y)$

$= \S  (E\Delta G\Delta H)\S  ((E\Delta E\Delta F\Delta G\Delta H) \cap X) \S  ((G\Delta G\Delta H\Delta E\Delta K)\cap Y)$

$= \S  (E\Delta G\Delta H) \S  ((E\Delta F\Delta E\Delta G\Delta H)\cap X) \S  ((G\Delta K\Delta E\Delta G\Delta H)\cap Y)$

$= \S  (E\Delta G\Delta H) \S  ((M\Delta E\Delta G\Delta H)\cap X) \S  ((T\Delta E\Delta G\Delta H)\cap Y)$

$= \S  (E\Delta G\Delta H) \S  (M\cap X)   \S  ((E\Delta G\Delta H)\cap X)\S  (T\cap Y) \S  ((E\Delta G\Delta H)\cap Y)$

$= \S  (M\cap X) \S  (T\cap Y) \S  (E\Delta G\Delta H) \S  ((E\Delta G\Delta H)\cap X) \S  ((E\Delta G\Delta H)\cap Y)$

$= \S  (M\cap X) \S  (T \cap Y) \S  ((E\Delta G\Delta H) \cap N) \S  (E\Delta G\Delta H)\cap X) \S  ((E\Delta G\Delta H) \cap Y)$

$=\S  (M\cap X) \S (T\cap Y) \S  ((E\Delta G\Delta H)\cap(N\Delta X\Delta Y))$

$= \S  (M\cap X) \S  (T \cap Y) \S  ( (E \Delta  G \Delta  H)\cap (X \Delta  Y)') $

\

In short:

\begin{equation}
	\S (A) \S (B\cap X) \S  (C\cap Y) = \S  (M\cap X) \S  (T \cap Y) \S  ( (E \Delta  G \Delta  H)\cap (X \Delta  Y)') 
\end{equation}

\

In these steps we have used some properties of set operations $\cup$, 
$\Delta$ , $\cap$, $\S$  : Union of disjoint sets coincides with the symmetric difference;
the operation $\Delta$  has a neutral element $\emptyset$ and the inverse of
any set is the same set itself; $\Delta$  is associative and commutative, i.e., its
order does not matter nor the parentheses; $\cap$ is distributive with
respect to $\cup$   or $\Delta$; $N$ is the neutral element of $\cap$ (See section 3). We also used property  (17) of parity.

\

Now, we can substitute (29) and (30) into (28) to get

\

$W(X,Y) = \sum    \S  (M\cap  X) \S (T\cap  Y) \S ((E\Delta G\Delta H) \cap (X \Delta  Y)') R (T\cup  I)$

\

To find the domains of the new variables, let us list all conditions we
have: $X<N$, $Y< N$ from (27) and $ A< N$, $B \cap C< A < B\cup  C$ from (28),
$I<(T\cup  M)'$ from (29), where $A= E\cup  G\cup  H$, $B = F\cup  G\cup  H$, $C = H\cup  E\cup  K$,
$M = F\cup  E$ and $T = G\cup  K$ as defined just after (28). The domains of
new variables can be defined from these conditions in multiple ways,
but we adopt one that will allow us great simplification: $T < N$
($T$ is taken as an independent variable, which can run over all subsets
of $N$), $M < T'$ (for $M\cap T = \emptyset$), $I < (T\cup  M)'$, $E< M$ (since $M= F\cup  E$),
$G < T$ (since $T = G\cup  K$), $H < (T\cup  M)' $ (for $H$ has a vacuum intersection
with $G\cup  K\cup  F\cup  E = T\cup  M )$. Reordering:

\begin{equation}
	W(X,Y) = \sum_{C1 }    \S  (M\cap X) \S (T\cap Y) R (T\cup  I) \sum_{C2} \S  ((E \Delta G\Delta H) \cap (X \Delta Y)')
\end{equation}

where $C1$ stands for $T < N$, $M < T'$, $I < (T\cup  M)'$ and $C2$ stands for $E<M$, 
$G< T$, $H< (T \cup M)'$.  As 

$(E \Delta G\Delta H) \cap    (X \Delta  Y)' = (E \cap    (X \Delta  Y)') \Delta  (G\cap   (X \Delta  Y)') \Delta (H \cap (X\Delta  Y)')$, 

and each of these three terms is disjoint
from the others, then by (15)

$\S  ( (E \Delta G\Delta H) \cap  (X\Delta Y)') = \S  (E \cap  (X \Delta  Y)')
\S  (G \cap (X \Delta  Y)') \S  (H \cap (X\Delta Y)')$.

 Therefore, the second term in (31 )
can be rearranged by (16) as

\

$\sum_{C2 } \S ((E \Delta G\Delta H) \cap    ( X \Delta  Y)')$

$= \sum_{E<M  } \sum_{G<T } \sum_{ H<(T\cup  M)'} \S  (E \cap (X\Delta Y)')    \S  (G \cap (X \Delta  Y) ')   \S  (H\cap  (X\Delta Y)')
$

$= \sum_{E<M  } \S  (E \cap (X\Delta Y)') \sum_{G<T }    \S  (G \cap (X \Delta  Y) ')\sum_{ H<(T\cup  M)'}    \S  (H\cap  (X\Delta Y)')
$
\

\

Applying ( 14) to each of these sums, we have that the  terms that do not vanish fulfill

\

 $M\cap  (X\Delta Y)' = \emptyset$; 
 
 $T \cap  (X\Delta Y)' = \emptyset $;      

$(T\cup  M)' \cap  (X\Delta Y)' =
(T \cup   M \cup   (X\Delta Y))' = \emptyset $.     

\

These three equalities hold if $X\Delta Y = N$, i.e., if
$Y = X'$. Replacing $E \Delta G\Delta H$ by $A$, we get:

\

$\sum_{C2}  \S  ( (E \Delta G\Delta H) \cap  (X \Delta  Y)') = \sum_{ A<N } \S  (A\cap (X\Delta Y)')= \sum_{ A<N } \S  (A\cap \emptyset )$

$\hspace{4.8cm}  = \sum_{ A<N } \S  (\emptyset ) = \sum_{ A<N } 1 =  2^n$ 

\

Let us replace this expression in (31)

$	W(X,Y) = \sum_{C1 }    \S  (M\cap X) \S (T\cap Y) R (T\cup  I) \sum_{C2} \S  ((E \Delta G\Delta H) \cap (X \Delta Y)')$

 and replacing also $Y$ by $X'$, we have that

\begin{equation}
	W (X,X') = 2^n \sum_{C1 } \S  (M \cap X) \S  (T \cap X') R(T \cup   I)
\end{equation}

where $C1$ stands for $T < N$, $M < T'$, $I < (T\cup  M)'$. Now let, $L = T\cup  I$,
then (32) becomes

\begin{center}
$W(X,X') = 2^n \sum_{C3 } \S  (M \cap X) \S  (T \cap X') R(L)$
\end{center}

where $C3$ determines the domains of the variables: 

$L< N$ (the independent
variable), 

$ T < L$ (for $L= T \cup   I)$, 

$M < L'$ (since, from $I <
(T\cup  M)'$ it ensues that $M\cap I = \emptyset $ or $M<I'$, AND, by construction, $M\cap  T= \emptyset$ or $M<T'$. Therefore, $M < T' \cap I'    = (T \cup I)' = L'$). Then

$$ W(X,X') = 2^n\sum_{L<N  }  R(L) \sum_{T<L } \S  (T \cap  X') \sum_{M<L' } \S(M\cap X)
  $$ 

The sum over M implies that $W (X,X')$ is nonzero if $L' \cap    X = (L\cup  X')'
= \emptyset$, then $L\cup  X' = N$, while the sum over $T$ implies that the only
important terms are those determined by $ L \cap    X' = \emptyset$. Together, this
is fitted if $L=X$. Then $\sum_{L<N  }  R(L)$ reduces to $R(X)$ and by (14)

\

$ W(X,X') =2^nR(X) \sum_{T<X  }    \S  (T\cap   X') \sum_{ M<X'}   \S  (M\cap   X) $

$\hspace{1.6cm} =2^n R(X) 2^{\#X } 2^{\#X' } = 2^n R(X) 2^{\#X +\#X' } = 2^n R(X) 2^{\#N} =2^{2n} R(X)$

\

Now, coming back to (27) we have, at last, the required identity (26'):

\

$t' (N) = 2^{-2n} \sum_{X<N }    \sum_{ X'<N }   2^{2n} R (X) t (X) t (X') = \sum_{X<N }    R(X) t (X) t (X')$
 
 \
 
where one sum is omitted because $X'$ is completely determined by $X$
and can take just one value.

\

Example 1: One locus: According to (26) the evolution of the transformed
gamete frequencies of one locus $t_a$ is given by

$ t'_a = \sum_{B<\{a\} } R (B/  \{a\} ) t (B) t (  \{a\} -B)$

$= R (\emptyset /  \{a\} ) t (\emptyset) t ( \{a\} ) + R( \{a\} / \{a\}) t ( \{a\} t (\emptyset))$

\begin{equation}
	t'_a = (R (\emptyset /  \{a\} ) + R (  \{a\} / \{a\} )) t (\emptyset) t ( \{a\} ) = t ( \{a\} ) = t_a 
\end{equation}

because, according to (9), $R (  \emptyset/  \{a\} ) + R ( \{a\} /  \{a\} ) = 1$ and $t (  \emptyset) = 1$
from (23) (this implies that $t' (\emptyset  ) = t (\emptyset )$). On the other hand, from
the inverse transform (21) we have that

$	p' ( \{a\} / \{a\} ) = 2^{-\# \{a\} } (\S  ( \{a\}  \cap  \emptyset )    t' (\emptyset) + \S  ( \{a\} \cap     \{a\} ) t' ( \{a\} ))$

$\hspace{1.9cm}  = 2^{-\# \{a\} } (\S  (   \emptyset )    t' (\emptyset) + \S  ( \{a\}  ) t' ( \{a\} ))$

$\hspace{1.9cm}  = 2^{-1}
	(1-t'_a) = 2^{-1}(1-t_a) = p ( \{a\} / \{a\} )$
	
Or 	

\begin{equation}
	p' ( \{a\} / \{a\} ) =  p ( \{a\} / \{a\} )
\end{equation}

where in the last equality we have recalled identity (25); similarly

$	p' ( \emptyset / \{a\} ) = 2^{-\# \{a\} } \S  ( \emptyset  \cap  \emptyset )    t' (\emptyset) + \S  ( \emptyset \cap     \{a\} ) t' ( \{a\} ))$

$\hspace{1.9cm}  = 2^{-1}
	(1+t'_a) = 2^{-1}(1+t_a) = p ( \emptyset / \{a\} )$
	
Or 	

\begin{center}
$$\hspace{4.4cm}	p' ( \emptyset / \{a\} ) =  p ( \emptyset / \{a\} ) \hspace{3.8cm} \hfill {(34')}$$
\end{center}

These equations say that the transformed frequency of one single
locus and its marginal gamete frequencies do not evolve in any way.
This is   simply the Hardy Weinberg law of equilibrium which is (34) with $n= 1$,  under the assumption that zygotes have just one locus.

\

Example 2: One simple calculation. Let $N= \{1,2\}$, $p ( \emptyset  ) = 3/10$, $p ( \{1 \}  ) = p (\{2\}) = 2/10$, $ p(N) = 3/10$. We use (8) to calculate the diverse values of

$$p (B/A) = \sum_{C<A'} p(B\cup C)$$

$p ( \emptyset /\emptyset ) = p(\emptyset) +  p ( \{1 \}  ) + p ( \{2 \}  ) +p( \{1,2\} )= 1$

$p ( \emptyset / \{1\} ) = p ( \emptyset    \cup   \emptyset   ) + p ( \emptyset    \cup    \{2\} ) = 3/ 10+ 2 /10 = 5/10 = 1 /2$

$p ( \{1\} / \{1\} ) = p ( \{1\} \cup   \emptyset ) + p ( \{1\} \cup   \{2\} )= 2/10 + 3/10 = 5/10=1/2$

Note that $p (\emptyset/ \{1\} ) + p ( \{1\} /  \{1\} ) = 1$.

$ p (\emptyset   /  \{2\} ) = p (\emptyset    \cup  \emptyset   ) + p (\emptyset   \cup    \{1\} ) = 3/10 + 2/10 = 5/10 = 1/2$.

$p ( \{2\} /  \{2\} ) = p ( \{2\} \cup   \emptyset) + p ( \{2\} \cup    \{1\} ) = 2/10 + 3/10=5/10= 1/2$.

$p (C/N) = p (C)$ for any $C$ according to (22).

The transformed frequencies are calculated according to (19):

$$	t(A) = \sum_{ X<A} \S(X) \sum_{C<A'}  p(X \cup C)  =\sum_{X<A}  \S  (X)p (X/A)$$

$t (\emptyset ) = 1$, as stated in (23 ).

$t ( \{1\} ) = \S  (\emptyset) p ((\emptyset /  \{1\} ) + \S  ( \{ 1\} ) p ( \{1\} / \{1\} ) = 1/2-1/2 = 0$.

$t ( \{2\} ) = \S    ( \emptyset) p (\emptyset) / \{2\} ) + \S  (  \{2\} ) p ( \{2\} / \{2\} ) = 1/2-1/2 = 0$.

$t ( \{1,2\} ) = \S  ( (\emptyset) p ( (\emptyset) / \{1,2\} ) 
+ \S  ( \{1\} p ( \{1 \} / \{1,2\} ) 
+ \S  ( \{2\} ) p ( \{2\} \{1,2\} )
+ \S  ( \{1,2\} p ( \{1,2\} /  \{1,2\} )= 3/10- 2/10 -2/10 +3/10 = 6/10-
 4/10 =1/5$.

Observe that $ t (\emptyset) + t ( \{1 \} ) + t ( \{2\} ) + t (  \{1,2\} ) \ne  1$. This means that,
until now, the only way to generate numbers that can serve as
transformed frequencies, i.e., that as a result of inverse transformation
(20) could restore positive numbers summing up to one, is to
transform ordinary frequencies according to the conventional definition
of transformed frequency ( 18).

Let us verify for some cases the inverse transform (21): given that $B<A$, then

$p (B/ A) = 2^{-\# A} \sum_{C <A}   \S  (B\cap C) t (C)$.

$p(\emptyset/\emptyset)=2^{-\# \emptyset} \S \emptyset t(\emptyset) =2^{-0}= 1$.

$ p(\emptyset/ \{1\} )=2^{ -\# \{1\}}  (\S \emptyset t(\emptyset ) + \S  \emptyset t( \{1\} )= 1/2(1 +0)= 1/2 $.

$p(\emptyset / \{1,2\} )=2^{ -\#  \{1,2\}} (\S \emptyset t(\emptyset) +\S \emptyset t( \{1\} )+\S  \emptyset t( \{2\} )+ \S \emptyset t ( \{1,2\} ))$

$\hspace{1.8cm}= 1/4 (1-0-0+1/5)= 1/4(6/5)=6/20=3/10$.

$p ( \{ 1\}  / \{1,2\} )=2^{ -\#  \{1,2\}} ( \S  ( \{ 1\} \cap   \emptyset) t(\emptyset) + \S  (  \{1\} \cap     \{1\} ) t (  \{1\} +\S  ( \{ 1\} \cap     \{2\} )
 t (  \{2\} )$
 
 $\hspace{2.4cm} + \S  (  \{1\} \cap     \{1,2\} ) t (  \{1,2\} )) = 1/4 ( 1-0+0-1/5) = 1/5.$
 
$p (  \{1,2\} / \{1,2\} )= 2^{ -\#  \{1,2\}} (\S  (  \{1,2\} \cap    \emptyset) t(\emptyset) + \S  (  \{1,2\} \cap     \{1\} t  \{1\} $

$\hspace{2.9cm} + \S  ( \{1,2\} \cap  \{2\} )t( \{2\} )+\S ( \{1,2\} \cap    \{1,2\} )t( \{1,2\} ))$

$\hspace{2.5cm} 
= 1/4 (1-0-0+ 1/5)= 1/4 (1+1/5)=6/20 = 3/10$;

\

Let $R (\emptyset   ) = R (  \{ 1,2\} ) = 1/3$ and $R (  \{1\} ) = R ( \{2\} ) = 1/6$; the
probability of no recombination is $R(\emptyset   ) + R ( \{1,2\} ) = 2/3$. The
marginal recombination distribution is defined for $B<A$ by   (6'):  

$$ R (B/A) = \sum_{C<A'}    R(B \cup   C)$$

$ R (\emptyset/\emptyset) =R (\emptyset) +R  \{1\} ) + R (  \{2\} ) + R ( \{1,2\} ) = 1$

$R (\emptyset   /( \{1\} ) = R (\emptyset   \cup  \emptyset   ) + R(\emptyset   \cup    \{2\} ) = 1/3 + 1/6= 1/2$;

$R ( \{1\} / \{1\} ) = R ( \{1\} \cup   \emptyset) + R ( \{1\} \cup   \{2\} ) = 1/6 + 1/3 = 1/2$;

We have that 

$R (\emptyset   /  \{1\} ) + R ( \{1\} / \{1\} )= 1.$

$R (\emptyset   /  \{2\} ) = R (\emptyset   \cup  \emptyset  ) + R (\emptyset     \cup    \{1\} )= 1/3 + 1/6 = 1/2$.

$R ( \{2\} / \{2\} )= R ( \{2\} \cup   \emptyset) + R ( \{2\} \cup   \{1\} )= 1/6 + 1/3 = 1/2$.

$R (C/N) = R (C)$ for any $C$ according to (22).

The transformed frequencies in the offspring generation are given by

$t' (A) = \sum_{B<A}  R (B/ A) t (B) t (A-B)$

$t' (\emptyset   ) = 1$

$t' ( \{1\} ) = R (\emptyset/  \{1\} ) t(\emptyset) t ( \{1\} - \emptyset) + R ( \{1\} / \{1\} ) t( \{1\} )t( \{1\} - \{1\} )$

$ \hspace{1.2cm}  = (1/2) (0) + (1/2) (0) = 0$;

$t' ( \{2\} ) = R (\emptyset/ \{2\} ) t (\emptyset) t ( \{2\} - \emptyset) + R ( \{2\} / \{2\} ) t ( \{2\}) t ( \{2\} - \{2\} ) = 0$

$t' (  \{1,2\} )= R (\emptyset/ \{ 1,2\} ) t(\emptyset) t( \{1 ,2\} -\emptyset ) +R( \{1\} / \{1,2\} )t ( \{1\} )t( \{1,2\} - \{1\} )$

$\hspace{1.8cm} 
+ R ( \{2\} / \{1,2\} )t( \{2\} )t( \{1,2\} - \{2\} )$

$\hspace{1.8cm} +R( \{1,2\} / \{1,2\} ) t( \{1,2\} )t( \{1,2\} - \{1,2\} )$

$\hspace{1.4cm} = (1/3) (1/5) + (1/6) (0) + (1/6) (0) + (1/3) (1/5) = 1/15+1/15 $

$\hspace{1.4cm}= 2/15 $

\

The diverse $p'$ of the offspring generation  can be calculated from the $t'$. Applying equation (21), we have 

$$	p' (B/A)= 2^{-\#A} \sum_{C<A}    \S  (B \cap   C) t'(C) \hspace{0.2cm} given \hspace{0.2cm} that \hspace{0.2cm} B<A	$$

$p'(\emptyset/\emptyset)=2^{-\# \emptyset} \S \emptyset t'(\emptyset) =2^{-0}= 1$.

\

$ p'(\emptyset/ \{1\} )=2^{ -\# \{1\}}  (\S \emptyset t'(\emptyset ) + \S  \emptyset t'( \{1\} )= 1/2(1 +0)= 1/2 $.

$ p'(\{1\}/ \{1\} )=2^{ -\# \{1\}}  (\S \emptyset t'(\emptyset ) + \S  \{1\} t'( \{1\} )= 1/2(1 -0)= 1/2 $.

\

$ p'(\emptyset/ \{2\} )=2^{ -\# \{1\}}  (\S \emptyset t'(\emptyset ) + \S  \emptyset t'( \{2\} )= 1/2(1 +0)= 1/2 $.

$ p'(\{2\}/ \{2\} )=2^{ -\# \{1\}}  (\S \emptyset t'(\emptyset ) + \S  \{2\} t'( \{2\} )= 1/2(1 -0)= 1/2 $.

\

$p'(\emptyset / \{1,2\} )=2^{ -\#  \{1,2\}} (\S \emptyset t'(\emptyset) +\S \emptyset t'( \{1\} )+\S  \emptyset t'( \{2\} )+ \S \emptyset t' ( \{1,2\} ))$

$\hspace{1.8cm}= 1/4 (1+0+0+2/5)= (1/4)(17/15) = 17/60$.

$p' ( \{ 1\}  / \{1,2\} )=2^{ -\#  \{1,2\}} ( \S  ( \{ 1\} \cap   \emptyset) t'(\emptyset) + \S  (  \{1\} \cap     \{1\} ) t' (  \{1\}$
 
 $\hspace{2.4cm} +\S  ( \{ 1\} \cap     \{2\} )
 t' (  \{2\} )  + \S  (  \{1\} \cap     \{1,2\} ) t' (  \{1,2\} )) $
 
 $\hspace{2.2cm}= 1/4 ( 1-0+0-2/15) = (1/4)(13/15) = 13/60.$

 $p' ( \{ 2\}  / \{1,2\} )=2^{ -\#  \{1,2\}} ( \S  ( \{ 2\} \cap   \emptyset) t'(\emptyset)+ \S  (  \{2\} \cap     \{1\} ) t' (  \{2\} $
 
 $\hspace{2.4cm} +\S  ( \{ 2\} \cap     \{2\} )
 t' (  \{2\} )  + \S  (  \{2\} \cap     \{1,2\} ) t' (  \{1,2\} )) $
 
 $\hspace{2.2cm}= 1/4 ( 1+0-0-2/15) = (1/4)(13/15) = 13/60.$

$p' (  \{1,2\} / \{1,2\} )= 2^{ -\#  \{1,2\}} (\S  (  \{1,2\} \cap    \emptyset) t'(\emptyset) + \S  (  \{1,2\} \cap     \{1\} t'  \{1\} $

$\hspace{2.9cm} + \S  ( \{1,2\} \cap  \{2\} )t'( \{2\} )+\S ( \{1,2\} \cap    \{1,2\} )t'( \{1,2\} ))$

$\hspace{2.5cm} 
= 1/4 (1-0-0+ 2/15)= 1/4 (1+2/15)=(1/4)(17/15)$

$ \hspace{2.5cm} = 17/60$;

Please, verify that marginal frequencies must add up to ONE.

\section{ FIXED POINTS OF GAMETE DYNAMICS}

We already noted, in (33) and (34), that the transformed frequencies
and the marginal frequencies relative to one locus are stable
notwithstanding reproduction with random mating. However, gamete
frequencies for other than one-locus would evolve to fulfill (26).
Would this evolution end in an equilibrium state, in a fixed point, or
would frequencies skip further and further? Let us begin to answer to this question by proving that  for each initial condition there is a given fixed point
which differs for different initial conditions and so, equilibriums are in general unstable. 

\

Using the ordinary symbol $\in$    as in $ a \in   A$ meaning that $a$ belongs to $A$, we define

\begin{equation}
	e_A= \prod_{a\in A }t_a
\end{equation}

where the letter $e$ is the first letter of equilibrium, for we are going to
prove that once the system arrives at $e_A$, for each $A$, it remains
invariable:

\begin{equation}
	e'_A = e_a 
\end{equation}
 
where $e'_A$ stands for the value of $e_A$ in the offspring generation. To
demonstrate (36) let us suppose that transformed frequencies $t (A)$
are equal to $e_A$ for each $A$. Let us substitute $e'_A$ by its value from (26):

$$ e'_A = \sum_{B<A} R(B/A)e_Be_{A-B} = \sum_{B<A} R(B/A) \prod_{b\in B }t_b \prod_{c\in A-B }t_c  
$$

$$ e'_A =   \sum_{B<A} R(B/A) \prod_{a\in A }t_a  =\prod_{a\in A }t_a  \sum_{B<A} R(B/A) = \prod_{a\in A }t_a = e_A
$$

where we have recalled that $B < A$ to make $(A-B) \cup   B$ = A, and the
fact that $R (B/ A)$ is a density function over $A$ whose integration over
the universal set $A$ renders 1. We have proved that $e_A$ defines indeed an equilibrium value. Let us relate its expression with frequencies. To that aim, let us now suppose that each $ t(A)$ equals $e_A$, then by applying
the inverse transform (20+ 21), we can calculate the equilibrium value of $p (A)$ for
any $A$:

\

$p(A) = p(A/N) = 2^{-n} \sum_{B<N} \S(A\cap B)e_B = 2^{-n} \sum_{B<N} \S(A\cap B)\prod_{a\in B }t_a $

\

To proceed further, let us learn from an example. Let $N= \{1,2,3,4\}$ and $A= \{1,3\}$,
then

$$\prod_{a\in A }(1-t_a) \prod_{b\in A' }(1+t_b)$$

$=(1-t_1) (1-t_3) (1+t_2)(1+t_4)$

$=(1-t_1) (1-t_3) (1+t_2+t_4+t_2 t_4)$

$=(1-t_3) (1+t_2+t_4+t_2 t_4 -t_1-t_1t_2-t_1t_4-t_1t_2t_4)$

$=( 1 +t_2 +t_4 +t_2 t_4 -t_1 -t_1 t_2 -t_1 t_4 -t_1 t_2 t_4 -t_3 -t_3 t_2 -t_3 t_4 -t_3 t_2 t_4 $

$\hspace{1cm}+t_3
t_1 +t_3 t_1 t_2 +t_3 t_1 t_4 +t_3 t_1 t_2 t_4 )$

$=1-t_1 +t_2 -t_3 +t_4 -t_1 t_2 +t_3 t_1 -t_1 t_4 -t_3 t_2 +t_2 t_4 -t_3 t_4 +t_3 t_1 t_2
-t_ 1 t_2 t_4 $

$\hspace{1cm} +t_ 3 t_1 t_4 -t_3 t_2 t_4 +t_3 t_1 t_2 t_4$

$= t(\emptyset) + \S  (A\cap     \{1\} ) t_1 + \S  (A\cap     \{2\} ) t_2 + \S  (A\cap     \{3\} ) t_3 + \S  (A\cap     \{4\} )
t_4$

$\hspace{1cm} + \S  (A\cap    \{1,2\} )t_1t_2+\S (A\cap    \{1,3\} )t_1t_3+\S (A\cap    \{1,4\} ) t_1t_4
+ \S  (A\cap     \{2,3\} ) t_2t_3 $

$\hspace{1cm} + \S  (A\cap     \{2,4\} ) t_2t_4+ \S  (A\cap     \{3,4\} ) t_3 t_4
+ \S  (A\cap     \{ 1,2,3\} ) t_1 t_2t_3 $

$\hspace{1cm} + \S  (A\cap     \{1,2,4\} ) t_1 t_2t_4 + \S  (A\cap     \{ 1,3,4\} )t_1t_3 t_4 + \S  (A\cap     \{2,3,4\} ) t_2t_3t_4 $

$\hspace{1cm}+ \S  (A\cap     \{ 1,2,3,4\} )   t_1 t_2 t_3t_4$

$=\sum_{B<N} \S (A\cap B)\prod_{a\in B }t_a$ where $t(\emptyset) = 1 $ and $\S(\emptyset)=1$.

Therefore, we have in  general:

 $$\sum_{B<N} \S (A\cap B)\prod_{a\in B }t_a = \prod_{a\in A }(1-t_a) \prod_{b\in A' }(1+t_b)$$
 
 Using this identity to rewrite $p(A)$, we get:

$p(A) = 2^{-n}  \prod_{a\in A }(1-t_a) \prod_{b\in A' }(1+t_b) $

$p(A)= \prod_{a\in A }(1-t_a)/2 \prod_{b\in A' }(1+t_b)/2$

where we have used the fact that $n = \#A+\#A'$ to introduce $2^n$ into the productores.

\

Recalling identity (25) we can write the equilibrium value of the frequencies:

\begin{equation}
	p(A) =\prod_{a\in A } p(\{a\}/\{a\}) \prod_{b\in A' } p(\emptyset/\{b\}) 
\end{equation}

This expression is simply the product of ordinary frequencies of
maternal alleles at loci in $A$ by the product of frequencies of paternal
alleles outside $A$.

Equation (37) shows that the system has a fixed point that depends on
the initial frequencies and does not depend on the scheme of recombination. For this reason, equilibriums are in general unstable. So, to relate this theory with experiment, it is mandatory to prove that the scientific research does not interfere with nature.

\section{DISEQUILIBRIUM MEASURES}

We have the expectancy that an initial condition would evolve toward  the equilibrium defined by it according to (37). To study this pretension, we need in first place to measure the difference between actual frequency $p (A)$ and its equilibrium value  given by  $\prod_{a\in N } p(\{a\}/\{a\}) \prod_{b\in A' } p(\emptyset/\{b\})$. That difference can be measured
directly by the function $p(A) -\prod_{a\in N } p(\{a\}/\{a\}) \prod_{b\in A' } p(\emptyset/\{b\})$. This is the most elementary and direct form to measure
gametic disequilibrium. But since $ p(A)$ and $\prod_{a\in N } p(\{a\}/\{a\}) \prod_{b\in A' } p(\emptyset/\{b\}) $ are unequivocally
determined by $t_A$ and $e_A$, then we define the gametic disequilibrium by :

\begin{equation}
	d_A=t_A-e_A
\end{equation}

Another measure of gametic disequilibrium is by means of the
Bennett measure of gametic disequilibrium $D_A$ defined by

\begin{equation}
	D_A= \sum_{B<A}\S  (A-B) p (\emptyset / B ) \prod_{a\in A-B }p  (\emptyset/   \{ a\} )
\end{equation}
  
These two forms to measure gametic disequilibrium are related
through the following identities:

\begin{equation}
	D_A= 2^{ -\#A}\sum_{B<A}\S  (A-B) d_Be_{A-B} =2^{ -\#A}\sum_{C<A}\S  (C) d_{A-C}e_ C
\end{equation}

\begin{equation}
	d_A= \sum_{B<A}\hspace{0.2cm} \sum_{C<B, C\ne \emptyset} 2^{ \#B}\S  (A-B) D_C \prod_{c\in B-C }p(\emptyset/\{c\})
\end{equation}

The two expressions of $D_A$ in (40) are equivalent and results from a change of variable: $C= A-B$, which, given that $B<A$, implies that $B =A-C$.

\

Let us prove (40). 

\

The proof for case $A = N$   follows. Equation (39) now reads:

\

$D_N= \sum_{B<N}\S  (N-B) p (\emptyset / B ) \prod_{a\in N-B }p  (\emptyset/   \{ a\} )$

$\hspace{0.7cm}  =\sum_{B<N}\S  (B') p (\emptyset / B ) \prod_{a\in B'}p  (\emptyset/   \{ a\} )   $

\

So, let us  replace in this equation the marginal probabilities by their transformed
frequencies in accordance with (21) and (25). Equation (21) reads:

\

$p (B/A)= 2^{-\#A} \sum_{C<A}    \S  (B \cap   C) t(C) $  given  that  $B<A$. Therefore:

$p (\emptyset/A)= 2^{-\#A} \sum_{C<A}    \S  (\emptyset \cap   C) t(C) $

$p (\emptyset/B)= 2^{-\#B} \sum_{C<B}    \S  (\emptyset \cap   C) t(C) $

\

while equation (25) is:

\

$	p (\emptyset / \{a\}) =  2^{-1} (\S \emptyset t(\emptyset)+ \S  (\emptyset)t(\{a\}))=(1/2)(1 +t_a)
$

\

Replacing these equations in the last version of (39), we get:

 \

$D_N   = \sum_{B<N} \S  (B') (2^{-\#  B}\sum_{C<B } \S  (\emptyset \cap C)t (C) ) \prod_{a\in B' } ( 1+ ta) /2$

$\hspace{0.7cm} = \sum_{B<N} 2^{-\#  B} \S  (B')  \sum_{C<B }  t (C)  2^{-\#  B'} \prod_{a\in B' } ( 1+ t_a)  $

\

because $\S  (\emptyset \cap C)= 1$ and inside the productore ( $\prod$) there are $\#B'$ terms divided by 2. As $2^{-\#  B}  2^{-\#  B'} = 2^{-n}$, we get:

\

$D_N = 2^{- n} \sum_{B<N}  \sum_{C<B } \S  (B')  t (C)   \prod_{a\in B' } ( 1+ t_a)  $

\

Let us notice now that:

\

$(1+t_1) (1+t_2)= (1+t_1+t_2+t_1 t_2)  $

$(1+t_1) (1+t_2)(1+ t_3) = (1+t_1+t_2+t_1 t_2) (1+t_3) $

$\hspace{3.6cm}= 1+t_1+t_2+t_1 t_2 + t_3+t_1t_3+t_2t_3+t_1 t_2t_3$
 
$\hspace{3.6cm}= 1+t_1+t_2+ t_3+t_1 t_2 +t_1t_3+t_2t_3+t_1 t_2t_3$

\

So, we have in general:

\

 $\prod_{a\in B'} (1+ t_a) = \sum_{A<B'} \prod_{a\in A} t_a = \sum_{A<B'} e_A$

\

Therefore

\

$D_N = 2^{-n}\sum_{B<N }\sum_{C<B }\S (B')t(C)\sum_{A <B' } e_A$

$D_N = 2^{-n}\sum_{B<N }\sum_{C<B }\sum_{A <B' } \S (B')t(C)e_A$

\stepcounter{figure}
\psset{unit=0.05cm}
\begin{center}
\begin{pspicture}(0,0)(130,90)
\rput{0}(30,60){\psellipse[](0,0)(10,-10)}
\rput{90}(37.5,60){\psellipse[](0,0)(25,-22.5)}
\rput{0}(105,35){\psellipse[](0,0)(15,-15)}
\rput(35,70){C}
\rput(55,80){B}
\rput(120,50){A}
\pspolygon[](10,90)(130,90)(130,10)(10,10)
\end{pspicture}

\bigskip
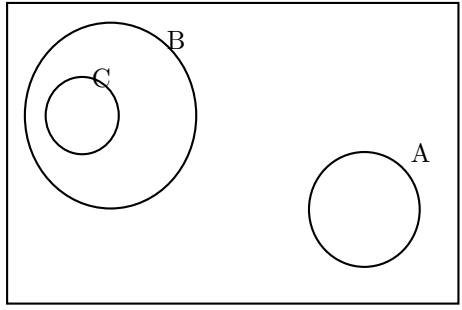
\textit{Figure \thefigure. This chart represents the situation given by $B<N$, $C<B$, $A< B'$, where $N$ is the universal set. The
change of variable  $K = B-C$ renders the equivalent set of conditions: $A<N$, $C<A'$, $K< (A\cup C)'$. } 

\end{center}

 Now, let $K=B -C$ (see figure 8), then     we can make a change of
variables with $A<N$, $C<A'$, $K< (A\cup C)'$. We have:

\

$D_N = 2^{-n}\sum_{A<N }\sum_{C<A' }\sum_{K <(A\cup C)' } \S (C\cup K)'t(C)e_A$

\

Let us prove now that $\S(C \cup K)' = \S C' \S K $. In effect, let us remember that $ K = B-C <C'$  and so $ K \cup C' = C'$, that parity is $\pm 1$ so its square is always 1, and then we apply property (15) of parity over disjoint sets:

 $ \S  ( (C\cup  K)')= 1\S  ( (C\cup  K)') = \S^2 K \S  ( (C\cup  K)') = \S K \S K \S  ( (C\cup  K)')  $
 
 $\hspace{2cm} = \S K   \S  ( K \cup (C\cup  K)')  = 
 \S K   \S  ( K \cup (C'\cap  K'))  $
 
 $\hspace{2cm} =  \S K   \S  ( (K \cup C') \cap (K \cup K')) =
 \S K   \S  (  C' \cap N) =  \S K   \S  (  C')$
 
 $\hspace{2cm} =   \S  (  C')\S K $

\

Therefore:

$D_N = 2^{-n}\sum_{A<N }\sum_{C<A' } \sum_{K<(A\cup C)' } \S (C')\S(K)t(C)e_A$

$D_N = 2^{-n}\sum_{A<N }\sum_{C<A' } \sum_{K<(A\cup C)' } \S (C')\S(K\cap N)t(C)e_A$

$D_N = 2^{-n}\sum_{A<N }\sum_{C<A' }  \S (C')t(C)e_A\sum_{K<(A\cup C)' } \S(K\cap N)$

\

Here the important terms in the last sum are determined by (14), a condition  that requires
that $N\cap    (A\cup  C)'= \emptyset$    i.e., that $A\cup  C=N$ which implies that $C=A'$, $C'=A$, while the variable $K$ runs over $K<(A\cup C)'= \emptyset$, so 
 $K=\emptyset$   while $\sum    \S (K\cap N)$ reduces to $ \S (\emptyset)  =1$. Therefore

\begin{equation}
	D_N =2^{-n}\sum_{A<N} \S (A)t(A')e_A 
\end{equation}
 
 Recalling that $ d_{A'} = t(A')- e_{A'}$, we get:

$D_N = 2^{-n} \sum_{A<N}  \S (A)(d_{A'} + e_{A'}) e_A$

$\hspace{0.65cm} = 2^{-n} \sum_{A<N}  \S (A)d_{A'} e_A +2^{-n} \sum_{A<N}  \S (A) e_{A'} e_A  $

If we use the fact that 
$e_{A'} e_A = e_N$ which can be factored in the second sum, we get:

$D_N = 2^{-n} \sum_{A<N}  \S (A)d_{A'} e_A +2^{-n} \sum_{A<N}  \S (A) e_{N}    $

$\hspace{0.65cm} = 2^{-n} \sum_{A<N}  \S (A)d_{A'} e_A +2^{-n}e_{N}  \sum_{A<N}  \S (A) = 2^{-n} \sum_{A<N}  \S (A)d_{A'} e_A    $

where we have applied (13)  to $N\ne \emptyset$
to render $\sum_{A<N}  \S (A)=0$. We end with

\

$D_N = 2^{-n} \sum_{A<N}  \S (A)d_{A'} e_A$

\

which is (40) for $A=N$. 

\

Now, let us engage in the proof of (41) for $A=N$:

\

	$d_N= \sum_{B<N}\hspace{0.2cm} \sum_{C<B, C\ne \emptyset} 2^{ \#B}\S  (N-B) D_C \prod_{c\in B-C }p(\emptyset/\{c\})$
 
 \
 
$\hspace{0.6cm}=R1 -R2$

\

where $R1$ differs from  $d_N$     in the terms involving $C=\emptyset$
and $R2$ is the corresponding compensation:

\

$R1 = \sum_{B<N}\hspace{0.2cm} \sum_{C<B} 2^{ \#B}\S  (B') D_C \prod_{c\in B-C }p(\emptyset/\{c\})$

\

$R2 = \sum_{B<N} 2^{ \#B}\S  (B') D_\emptyset \prod_{c\in B }p(\emptyset/\{c\})$

\

To prove (41) it is enough to show that $R1=t (N)$ and $R2=e_N$ because
by definition (38) $d_N = t(N) - e_N$.

\

Let us replace $D_C $ in $R 1$ by its value induced from (42)
above and renumber in the inverse order:

\

 $D_C = 2^{ -\#C}\sum_{K<C} \S(K) t(C-K) e_K = 2^{ -\#C}\sum_{K<C} \S(C-K) t(K) e_{C-K} $
 
$$R1 = \sum_{B<N}\hspace{0.2cm} \sum_{C<B} 2^{ \#B}\S  (B') [2^{ -\#C}\sum_{K<C} \S(C-K) t(K) e_{C-K} ] \prod_{c\in B-C }p(\emptyset/\{c\})$$

$$R1 = \sum_{B<N}\hspace{0.2cm} \sum_{C<B}\sum_{K<C} 2^{ \#(B-C)}\S  (B')   \S(C-K) t(K) e_{C-K}  \prod_{c\in B-C }(1+t_c)/2$$

where we have recalled (25) to replace the marginal probabilities. The last (1/2) can be factored out of the productore:

$$R1 = \sum_{B<N}\hspace{0.2cm} \sum_{C<B}\sum_{K<C} 2^{ \#(B-C)}\S  (B')   \S(C-K) t(K) e_{C-K} 2^{-\#(B-c) } \prod_{c\in B-C }(1+t_c) $$

The productore can be expanded using a formula   that was found  above:

$\prod_{a\in B'} (1+ t_a) = \sum_{A<B'} \prod_{a\in A} t_a = \sum_{A<B'} e_A$

or  

$\prod_{a\in M} (1+ t_a) = \sum_{A<M} \prod_{a\in A} t_a = \sum_{A<M} e_A$

If we apply this to $R1$ with $M = B-C$:

$$R1 = \sum_{B<N}\hspace{0.2cm} \sum_{C<B}\sum_{K<C}  2^{ \#(B-C)}  2^{ -\#(B-C)}\S  (B')   \S(C-K) t(K) e_{C-K}\sum_{O<B-C} e_O$$

$$R1 = \sum_{B<N}\hspace{0.2cm} \sum_{C<B}\sum_{K<C} \sum_{O<B-C} 2^{ \#(B-C)}  2^{ -\#(B-C)}\S  (B')   \S(C-K) t(K) e_{C-K} e_O$$

$$R1 = \sum_{B<N}\hspace{0.2cm} \sum_{C<B}\sum_{K<C} \sum_{O<B-C} \S  (B')   \S(C-K) t(K) e_{C-K} e_O$$

\

To realize what kind of situation we have, let us pay attention to
figure 9. 

\

Since $ B'$ and $C-K$ are disjoint sets, then by (15) 

$\S  (B') \S  (C-K)=
\S (B'\Delta (C-K) ) = \S (N\Delta B\Delta C\Delta K)$ 

 because $B' = N\Delta B$ and $C-K = C\Delta K$ since $K<C$.

On the other hand, $e_{C-K} e_O =e_{C\Delta K } e_O =e_{C\Delta K \Delta O}$.

\

Now, we can make a change of variables: let $X=B-(C \cup   O)$ and
$Y=C-K$; therefore $B=C \cup   O \cup   X=C\Delta O \Delta X$ and $C=K \cup   Y = K\Delta Y$.
Similarly $N\Delta B\Delta C\Delta K=N\Delta C\Delta O\Delta X\Delta C\Delta K=N\Delta O\Delta X\Delta K$ and $C\Delta K\Delta O =
K \Delta  Y \Delta  K \Delta  O = Y \Delta  O$ and $B-C=X \cup   O=X\Delta O$. We have at last that
$\S  (B') \S  (C-K) = \S  (N\Delta O\Delta X\Delta K) = \S  (N\Delta O\Delta K) \S  (X)$  and $e_{C-K} e_O=  e_{Y\Delta O}$.  

\

In short, the change of variables is:

$K$ is independent variable.

$O<K'$

$Y=C-K$; 

$X=B-(C \cup   O)$

$\S  (B') \S  (C-K) = \S  (N\Delta O\Delta X\Delta K) = \S  (N\Delta O\Delta K) \S  (X)$

$e_{C-K} e_O=  e_{Y\Delta O}$

\

Let us notice that   

 $R1 = \sum_{B<N}\hspace{0.2cm} \sum_{C<B}\sum_{K<C} \sum_{O<B-C} \S  (B')   \S(C-K) t(K) e_{C-K} e_O$ 

has the general form

 $R1 = \sum  \S  (B')   \S(C-K) t(K) e_{C-K} e_O$ 
 
 and with the change of variables, this expression takes the form
 
 $R1= \sum     \S  (N\Delta O\Delta K) \S  (X) t (K) e_{Y\Delta O}$

\stepcounter{figure}
\psset{unit=0.09cm}
\begin{center}
\begin{pspicture}(0,0)(125.94,126.24)
\rput(105,30){}
\rput(118.12,108.12){N}
\rput(11.25,98.12){}
\rput{90}(70.09,43.12){\psellipse[](0,0)(10,-9.48)}
\rput{90}(36.91,63.12){\psellipse[](0,0)(10,-9.48)}
\rput{90}(41.65,65.62){\psellipse[](0,0)(22.5,-18.96)}
\rput{90}(51.13,63.12){\psellipse[](0,0)(40,-37.92)}
\rput(79.57,53.12){O}
\rput(46.39,73.12){K}
\rput(55.87,88.12){C}
\rput(70.09,68.12){}
\rput(79.57,98.12){B}
\pspolygon[](7.81,5.31)(125.94,5.94)(125.18,126.24)(7.06,125.62)
\psline(29.8,83.12)(53.5,83.12)
\psline(23.88,73.12)(59.43,73.12)
\psline(22.7,63.12)(27.43,63.12)
\psline(46.39,63.12)(60.61,63.12)
\psline(26.25,53.12)(57.06,53.12)
\psline(30.39,48.12)(52.91,48.12)
\psline(23.88,58.12)(28.62,58.12)
\psline(44.61,58.12)(59.43,58.12)
\psline(22.7,68.12)(28.62,68.12)
\psline(44.61,68.12)(60.61,68.12)
\psline(26.25,78.12)(57.65,78.12)
\rput(62.39,86.24){*}
\rput(73.05,68.12){*}
\rput(19.74,80){*}
\rput(15.59,60.62){*}
\rput(29.22,36.88){*}
\rput(77.79,81.24){*}
\rput(30.4,88.75){*}
\rput(23.28,48.75){*}
\rput(63.57,75.62){*}
\rput(86.09,63.74){*}
\rput(36.33,91.24){*}
\rput(83.12,69.38){*}
\rput(25.65,83.74){*}
\rput(20.91,75.62){*}
\rput(25.65,42.5){*}
\rput(19.14,56.25){*}
\rput(17.96,66.24){*}
\rput(34.54,39.38){*}
\rput(36.33,35.62){*}
\rput(44.61,35.62){*}
\rput(42.84,28.12){*}
\rput(57.65,30){*}
\rput(65.35,30){*}
\rput(52.91,31.25){*}
\rput(49.35,39.38){*}
\rput(54.09,43.12){*}
\rput(58.83,35.62){*}
\rput(70.09,57.5){*}
\rput(82.53,45){*}
\rput(60.02,50.62){*}
\rput(78.98,63.12){*}
\rput(74.24,73.74){*}
\rput(67.13,65){*}
\rput(67.13,81.24){*}
\rput(71.87,85){*}
\rput(60.61,95){*}
\rput(68.31,92.5){*}
\rput(51.13,96.88){*}
\rput(48.17,94.38){*}
\rput(45.21,96.88){*}
\rput(39.28,95){*}
\rput(16.77,79.38){}
\end{pspicture}

\bigskip
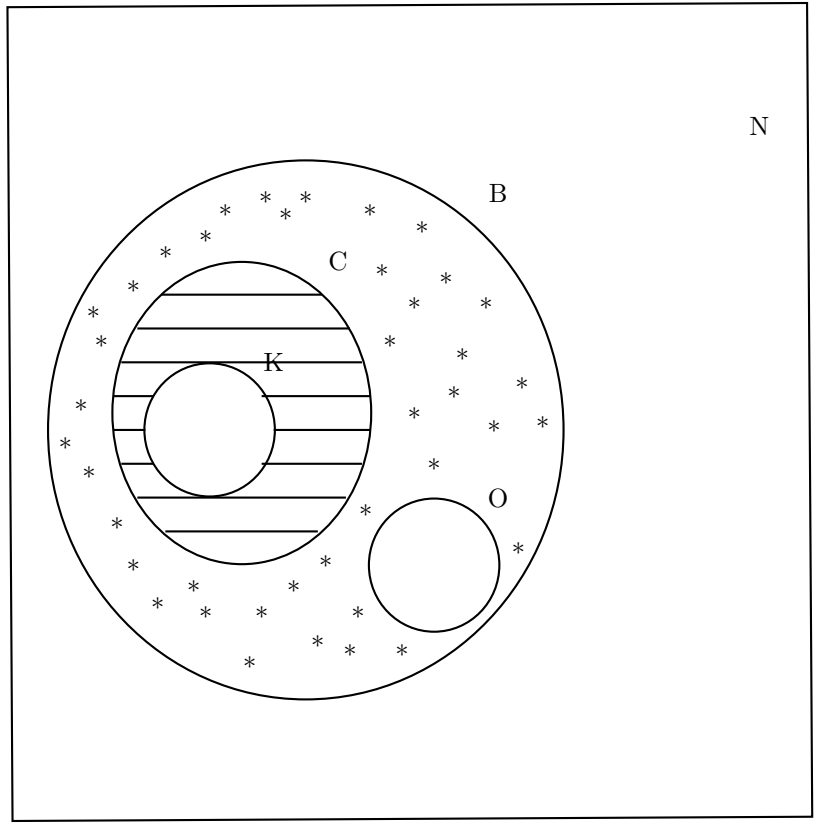
\textit{Figure \thefigure. This is a graphic representation of the following conditions: $B<N$, 
$C <B$, $ K<C$, $O <   B-C$. New disjoint sets $X=B -(C\cup  O)$ (in stars) and $ Y= C-K$ (in lines) are indicated. } 

\end{center}

Therefore, the domain of the sums can be defined by  

$K < N$,
$O<K'$, $Y< (K\cup  O)'$, $X < (K\cup  O\cup  Y)'=N\cap   (K\cup  O\cup  Y)'$, 
 
This set of conditions will be referred to as $F$ and we can rewrite

$R1=
\sum_F    \S  (N\Delta O\Delta K) \S  (X) t (K) e_{Y\Delta O}$

Let us separate the terms with $\S(X)$:

\

$R1 = \sum     \S  (N\Delta O\Delta K)  t (K) e_{Y\Delta O}  \sum_{ X < (K\cup  O\cup  Y)'} \S  (X)  $

\

Let us apply the   law (13) to $\S  (X)$: when $X 
<  (K \cup   O \cup   Y)'$, the only nonzero terms that go in this sum are
those corresponding to the case $  (K \cup   O \cup   Y)' = \emptyset$, i.e., $X = \emptyset$,   rendering that $\sum \S  (X) = 1$. But if $  (K \cup   O \cup   Y)' = \emptyset$ then $K \cup   O \cup  
Y=N$   and given that these three sets are
disjoint this is equivalent to $K'=   Y\cup  O = Y \Delta O$. Therefore

\

$R1 = \sum     \S  (N\Delta O\Delta K)  t (K) e_{K'}  $

\

Hence condition $F$
reduces to $K <N, O<K'$. So,

\

 $R1 = \sum_{K<N  } \sum_{  O<K' }  \hspace{0.2cm}  \S (N\Delta O\Delta K)  t(K)e_{K'}  $
 
 \
 
 Let us notice now that
 
 $ \S (N\Delta O \Delta K) = \S( (O \cup K)')  = \S( O' \cap K')  = \S O \S O \S( O' \cap K')  $
 
 $ \hspace{2cm} = \S O \S ( O \cup (O'\cap K') ) = \S O \S ( (O \cup O') \cap  (O \cup K') ) $
 
 $  \hspace{2cm} = \S O \S (N \cap K') = \S O \S K' = \S K' \S O $

 Therefore:
 
 \
 
 $ R1 =  \sum_{K<N}  \sum_{O<K' }     \S  (K') \S  (O)t(K)e_{K'}   $
 
 $\hspace{0.6cm}  =  \sum_{K<N}     \S  (K') t(K)e_{K'}  \sum_{O<K' }  \S  (O) $ 

\

which gives after (13) that the important terms in the last sum are when $K'=\emptyset$ . i.e., $K=N$,
then $\S  (K')= \S  (\emptyset)= 1$ and $e_{K' }=e_{\emptyset }=1$. We are left with $t(N)$ alone and we have proved, as it was promised, that

\begin{equation}
	R1 =\sum_{B<N } \sum_{C<B} 2^{\#B} \S  (B') D_C \prod_{c \in B-C}  p(\emptyset/ \{c\} )= t(N)  
\end{equation}
 
\

Now, let us retake $R2$, which must be proved to equal $e_N$:

\

$R2 = \sum_{B<N } 2^{\#B} \S  (B') D_\emptyset \prod_{c\in B}p(\emptyset/\{c\}) =  \sum_{B<N } 2^{\#B} \S  (B') D_\emptyset 2^{-\#B}\sum_{C< B}e_c $

$\hspace{0,6cm}=  \sum_{B<N } \sum_{C<B } \S(B') e_C$

\

where the productore of the marginal distributions has been expanded, as usual, into a sum of $e_C$ and
we have used the fact that $ D_\emptyset= 1$. 
Let $B-C=Z$ be a new variable. Then 

$B = C \cup   Z$

$B'=(C \cup   Z)'$

and, as before, 

$\S  (B')= \S (C') \S  ( Z)$

\

With this change of variable, $R2$ becomes:

$R2 = \sum_{C<N }\sum_{Z<C' }\S  ( (C\cup  Z)'\} e_C = \sum_{C<N} e_C  \S  (C')\sum_{Z<C'}  \S  (Z)$

according to ( 13) in the last sum there is only one nonzero term
corresponding to $C'=\emptyset$ , i.e., $C=N$ and therefore $R2=e_N\S(\emptyset) = e_N$. We have proved that:

\begin{equation}
	R2 = \sum_{B<N } 2^{\#B} \S  (B') D_\emptyset \prod_{c\in B}p(\emptyset/\{c\})  = \sum_{B<N } \sum_{C<B }   \S  (B') e_C = e_N  
\end{equation}

This ends the proof that the two measures of disequilibrium $ d$ and $D$ are related by (40+41).
 
 \
 
One of the most interesting properties of the disequilibrium value $D_A$
defined by (39) is that it obeys the same recurrence equation as the
transformed frequencies. If $D'_A$ is the gamete disequilibrium in the
offspring generation, then

\begin{equation}
	D'_A =   \sum_{B<A} R(B/A) D_B   D_{A-B}
\end{equation}

Thanks to this equation, one algorithm can be used to calculate both
the evolution of transformed frequencies and that of gamete disequilibrum.

\

To demonstrate (45) we express $D'_A$    in terms of transformed
frequencies in the offspring generation, as stated in (42), reorder, and then
replace these transformed frequencies by their values given by
recurrence equation (26):

\

$D'_N = 2^{-n } \sum_{A<N} \S (A)t' (A') e_A = 2^{-n} \sum_{A<N} \S  (A')t' (A)e_{A'}$

$\hspace{0.6cm}= 2^{-n} \sum_{A<N} \S  (A') e_{A'}\sum_{B<A}R(B/A)t(B)t(A-B)$

\

Recalling (6') to replace $R(B/A)$, we get:

\begin{equation}
	D'_N = 2^{-n } \sum_{A<N} \sum_{B<A}\sum_{C<A'} \S  (A') e_{A'}R(B\cup  C) t(B)t(A-B) 
\end{equation}

Now we need to express $t(B)$ and $t(A-B)$ in terms of the $D_k 's$, which
is done by inducing their values from $t(N)$ in ( 43) :

\begin{equation}
	t(A) = \sum_{B<A}\hspace{0.2cm} \sum_{C<B} 2^{\#B }\S  (A-B)D_C\prod_{c\in B-C} p(\emptyset/ \{c\})
\end{equation}

$t(A) = \sum_{B<A}\sum_{C<B} 2^{\#B }\S  (A-B)D_C   2^{-\#(B-C) } \sum_{K<B-C} e_K$

$\hspace{0.8cm}= \sum_{B<A}\sum_{C<B}\sum_{K<B-C} 2^{\#C }\S  (A- B)D_C     e_K$

\

Please, make a graphic of the conditions above: $B<A$, $C<B$, $K <B-C$  taking $A$ as the universal set and verify the next change of variables:

\

Let $ Y = B-(C\cup  K)$ or $B= Y\cup  C\cup  K $ and $ A-B = A- (C\cup  Y\cup  K)$ which
renders for $B<A$ that $\S  (A - B)= \S  (A-(C\cup  Y\cup  K) ) =  \S  ( (C\cup  Y\cup  K)' ) $, where the complement is relative to $A$. We have, as it was done before:

$ \S  ( (C\cup  Y\cup  K)' ) =  \S  ( (C\cup    K \cup Y)' ) = \S          (C\cup    K)' \S  Y   = \S  ( A-(C\cup    K) ) \S  Y$

Therefore:

$\S  (A - B)= \S  ( A-(C\cup    K) ) \S  Y$ 

  On the other hand, the set of conditions $B<A$, $C<B$, $K <B-C$ is equivalent to
$C<A$, $K<A-C$, $Y <A -(C \cup   K)$. Replacing this in $t(A)$ above, we get:

\

$t(A) = \sum_{C<A} \sum_{K<A-C} 2^{\#C} D_C e_K \S(A- ( C \cup K) \sum_{Y<A-(C\cup K)} \S(Y)$

\
 
The last sum is ONE and  corresponds to the case in which $A-
(C \cup   K)= \emptyset$, i.e., $A = C \cup   K$ or $K = A-C$ and because
$\S  ( A-(C\cup    K) ) = \S  (\emptyset   ) = 1$, finally we get :

\begin{equation}
	 t(A) = \sum_{C<A} 2^{\#C} D_C e_{A-C}
\end{equation}

Therefore

\

$t(B) t (A-B) = (\sum_{X<B}  2^{\#X} D_X e_{B-X} ) (\sum_{Y<A-B}  2^{\#Y} D_Y e_{(A-B)-Y} ) $

$\hspace{1.8cm } = \sum_{X<B}  \sum_{Y<A-B} 2^{\#(X\cup Y)} D_XD_Y e_{B-X}    e_{(A-B)-Y}  $

\

and coming back to (46)

\

$D'_N = 2^{-n} \sum_{C4} 2^{\#(X\cup Y)} D_XD_Y R(B\cup C)e_{A'}e_{B-X}    e_{(A-B)-Y}  \S(A')$
 
 \

where $C4$ stands for the condition given by $A<N$, $B<A$, $C<A'$,
$X<B$, $Y<A-B$, a situation that is visualized in figure 10.

Since $A'$,
$B-X$ and $(A-B)- Y$ are disjoint sets then we have that 

$e_{A'} e_{B -X}
e_{(A-B)-Y }= e_{A' \cup  (B-X)\cup  ((A-B)-Y)} =e_{(Y\cup X)'}$ 

 which is a more tractable
expression. 

\stepcounter{figure}
\psset{unit=0.05cm}
\begin{center}

\begin{pspicture}(0,0)(120,100)
\rput{0}(100,20){\psellipse[](0,0)(10,-10)}
\rput{0}(65,45){\psellipse[](0,0)(5,-5)}
\rput{0}(25,55){\psellipse[](0,0)(5,-5)}
\rput{0}(27.5,52.5){\psellipse[](0,0)(12.5,-12.5)}
\rput{90}(42.5,55){\psellipse[](0,0)(35,-32.5)}
\pspolygon[](0,0)(120,0)(120,100)(0,100)
\rput(30,60){X}
\rput(40,70){B}
\rput(60,90){A}
\rput(65,55){Y}
\rput(110,35){C}
\rput(110,95){N}
\rput(100,70){}
\psline(25.62,85)(58.12,85)
\psline(20,80)(65,80)
\psline(15.62,75)(69.38,75)
\psline(13.12,70)(71.88,70)
\psline(11.88,65)(73.12,65)
\psline(10,60)(18.12,60)
\psline(37.5,60)(75,60)
\psline(10,55)(15,55)
\psline(40,55)(75,55)
\psline(10,50)(15,50)
\psline(40,50)(75,50)
\psline(11.88,45)(16.88,45)
\psline(36.88,45)(60,45)
\psline(70,45)(73.12,45)
\psline(13.12,40)(71.88,40)
\psline(16.25,35)(68.12,35)
\psline(20,30)(65,30)
\psline(26.25,25)(58.75,25)
\psline(20,60)(25,65)
\psline(15,55)(20,60)
\psline(15,50)(20,55)
\psline(25,60)(30,65)
\psline(16.88,46.25)(21.88,51.25)
\psline(28.75,58.75)(33.12,63.12)
\psline(18.75,43.75)(25,50)
\psline(30,55)(36.25,61.25)
\psline(21.25,41.25)(38.75,57.5)
\psline(25,40)(40,55)
\psline(30,40)(40,50)
\end{pspicture}

\bigskip
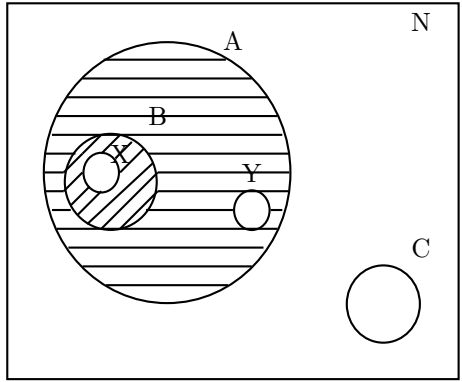
\textit{Figure \thefigure. Here we have the representation of: $A<N$, $B<A$, $C<A'$, $X<B$, $Y<A-B$. Subsets $E = B-X$ (oblique lines) and $F= A-(X\cup Y \cup E)$ (horizontal lines) are indicated. } 

\end{center}

\

To introduce a change of variables, let $E = B-X$, $F=
A- (X \cup   Y \cup   E)$, so that $A=X \cup   Y \cup   E \cup   F$. Therefore $\S  (A') = \S  ( (X \cup   Y \cup   E \cup   F )')  = \S  (X \cup   Y \cup   E )' \S  (F)$ and \\
$R (B \cup   C) =R(X \cup   E \cup   C)$. Hence, the old expression

$D'_N = 2^{-n} \sum_{C4} 2^{\#(X\cup Y)} D_XD_Y R(B\cup C)e_{A'}e_{B-X}    e_{(A-B)-Y}  \S(A')$

becomes

$$D'_N = 2^{-n} \sum_{C5} 2^{\#(X\cup Y)} D_XD_Y R(X\cup E \cup C) e_{(X\cup Y)'}    \S( (X \cup   Y \cup   E   )'  ) \S(F)$$

where $C5$ is given by $X <N$, $Y<X'$, $E < (X\cup  Y)'$, $C<(X\cup  E\cup  Y)'$, 
$F< (X\cup  E\cup  Y\cup  C)'$. We can separate the sum of $\S  (F)$   with the condition $F<(X\cup  E\cup  Y\cup  C)'$.
This sum matters when  $(X\cup  E\cup  Y\cup  C)' = \emptyset$, i.e., $X\cup  E\cup  Y\cup  C = N$ , or $C =
(X\cup  E\cup  Y)'$ and

$F<(X\cup  E\cup  Y\cup  C)' = ((X\cup  E\cup  Y) \cup  (X\cup  E\cup  Y)' )'  = N' = \emptyset $. 

In short, $F=\emptyset$ and the corresponding sum adds up to 1.

Then

$R (X\cup  E\cup  C) = R (X\cup  E\cup   (X\cup  E\cup  Y)') =
R (X\Delta E\Delta  N\Delta X\Delta E\Delta Y) $

$\hspace{2.3cm} = R (N\Delta Y)= R(Y')$

\

$D'_N = 2^{-n} \sum_{C6} 2^{\#(X\cup Y)} D_XD_Y R(Y')e_{(X\cup Y)'}    \S( (X \cup   Y \cup   E   )'  ) $

\

where $C6$ stands for $X<N$, $Y<X'$, $E< (X\cup  Y)'$, $C<(X\cup  E\cup  Y)'$ and using and old trick $\S( (X \cup   Y \cup   E   )'  )  =  
\S( (X \cup   Y )'  ) ) \S  (E)$. We note that
$\sum  \S  (E)$ can be factored with the condition that $E < (X\cup  Y)'$ rendering that $(X\cup  Y)'
= \emptyset$ or $X\cup  Y = N$,  i.e. $Y = X'$. We have

\

$D'_N = 2^{-n} \sum_{X<N}   2^{\#(X\cup  Y)} D_X D_{X'} R (X) e_\emptyset \S  ( (X\cup X')')$

 \
 
Using $\# (X\cup  X') = \# (N) = n $ and $e_\emptyset    = 1$ while $\S  (( X\cup X')') = \S (N') = \S  (\emptyset   ) = 1$, we get at last:

\

$D'_N =  \sum_{X<N}   D_X D_{X'} R (X) $

\

This finishes the proof of ( 45).

\

Example. Let us calculate some instances of the initial and subsequent disequilibrium given by (39) and (45) respectively. Formula (39) for the initial disequilibrium reads:

$$	D_A= \sum_{B<A}\S  (A-B) p (\emptyset / B ) \prod_{a\in A-B }p  (\emptyset/   \{ a\} )$$

We have:

\begin{enumerate}
	\item  $D_\emptyset = \S  (\emptyset   ) p (\emptyset    / \emptyset   ) = 1$, where we have making $\prod_{a\in \emptyset}p  (\emptyset/   \{ a\} ) =p(\emptyset/\emptyset)=1$
	
	\item  $D_{\{a\} } = \S  ( \{a\} -\emptyset   ) p (\emptyset    / \emptyset   ) p (\emptyset    /  \{a\} )+ \S  ( \{a\} - \{a\} ) p (\emptyset / \{a\} )
p (\emptyset/\emptyset) $

$\hspace{0.9cm} = \S  ( \{a\} ) p (\emptyset/ \{a\} ) + \S  (\emptyset ) p (\emptyset / \{a\} ) =
-p ( \emptyset    /  \{a\} ) + p ( \emptyset    /  \{a\} ) = 0 $

\item $ D_{ \{a,b\}} = \sum_{ B < \{a.b\}}   \S  ( \{a,b\}  -B) p (\emptyset   /B) \prod_{a\in (a,b)-B } p (\emptyset    / \{a\} )$
  
$\hspace{1.1cm}=\S ( \{a,b\} -\emptyset) p ( \emptyset    / \emptyset   ) p (\emptyset    /  \{a\} ) p ( \emptyset    /  \{b\} )+ \S  (  \{a,b\} -  \{a\} ) p (\emptyset/  \{a\}  )
p (\emptyset    /  \{b\} ) $

$\hspace{1.6cm} + \S  ( \{a,b\} -  \{b\} ) p (\emptyset/ \{b\} ) p (\emptyset/ \{a\} ) + \S ( \{a,b\} -  \{a,b\} )
p (\emptyset    / \{a,b\} ) p (\emptyset    / \emptyset   )$

$\hspace{1.1cm}=
p (\emptyset    /  \{a\} ) p (\emptyset    /  \{b\} )-p (\emptyset    /  \{a\} ) p (\emptyset    / \{b\} )- p (\emptyset    /  \{b\} ) p (\emptyset    /  \{a\} )$

$\hspace{1.6cm} + p (\emptyset    /  \{a,b\} )$

$\hspace{1.1cm}= p (\emptyset/ \{a,b\} ) -p (\emptyset/ \{b\} ) p (\emptyset/ \{a\} ). $

\item $D_{ \{a,b,c\}}  = \sum_{ B<A}    \S  ( \{a,b,c\}  -B) p (\emptyset    / B) \prod_{[a\in \{a,b,c\}-B]} p (\emptyset    /  \{a\} )$
  
$\hspace{1.3cm} = \S  ( \{a,b,c\} -\emptyset)p(\emptyset   /\emptyset   )p(\emptyset   / \{a\} )p(\emptyset   / \{b\} )p(\emptyset /\{c\})$

$\hspace{1.7cm}  
+ \S  ( \{a,b,c\} - \{a\} ) p (\emptyset/ \{a\} ) p (\emptyset/ \{b\} ) p(\emptyset/ \{c\} ) $

$\hspace{1.7cm} 
+ \S  ( \{a,b,c\} - \{b\} ) p (\emptyset/ \{b\} ) p (\emptyset/ \{a\} ) p (\emptyset/ \{c\} )$

$\hspace{1.7cm} + \S  (  \{a,b,c\} - \{c\} ) p (\emptyset /  \{c\} ) p (\emptyset /  \{a\} ) p (\emptyset /  \{b\} )$

$\hspace{1.7cm} + \S  (  \{a,b,c\} -  \{a,b\} ) p (\emptyset /  \{a,b\} ) p (\emptyset /  \{c\} )$

$\hspace{1.7cm} 
+ \S  ( \{a,b,c\} - \{a,c\} ) p (\emptyset/ \{a,c\} ) p (\emptyset/ \{b\} )$

$\hspace{1.7cm} 
+ \S  (  \{a,b,c\} - \{b,c\} ) p (\emptyset /  \{b,c\} ) p (\emptyset /  \{a\} )$

$\hspace{1.7cm} + \S  ( \{a,b,c\} - \{a,b,c\} ) p (\emptyset/ \{a,b,c\} ) p (\emptyset/\emptyset) $

\

$\hspace{1.3cm} =   
-p (\emptyset / \{a\} )p (\emptyset / \{b\} )p (\emptyset / \{c\} )$

$\hspace{1.7cm}+ p (\emptyset / \{a\} )p (\emptyset / \{b\} )p ( \emptyset / \{c\} )$

$\hspace{1.7cm}
+ p ( \emptyset /  \{b\} )p (\emptyset /  \{a\} ) p (\emptyset /  \{c\} )$

$\hspace{1.7cm}+ p (\emptyset / \{c\} )p (\emptyset /  \{a\} )p ( \emptyset /  \{b\} )$

$\hspace{1.7cm}-p (\emptyset / \{a,b\} )p (\emptyset / \{c\} )$

$\hspace{1.7cm}-p (\emptyset / \{a,c\} )p (\emptyset / \{b\} )$

$\hspace{1.7cm}-p (\emptyset /  \{b,c\} )
p (\emptyset /  \{a\} ) $

$\hspace{1.7cm}+ p ( (\emptyset) /  \{a,b,c\} )$

\

$\hspace{1.3cm} =  p (\emptyset/ \{a,b,c\}  - p (\emptyset/ \{a\} ) p (\emptyset/ \{b\} ) p (\emptyset/ \{c\} )$

$\hspace{1.7cm}
+ ( -p (\emptyset /  \{a,b\} ) p (\emptyset /  \{c\} ) + p (\emptyset /  \{a\} ) p (\emptyset /  \{b\} ) p (\emptyset /  \{c\} ))$

$\hspace{1.7cm}+ (-p (\emptyset/ \{a,c\} ) p (\emptyset/ \{b\} ) + p (\emptyset/ \{b\} ) p (\emptyset/ \{a\} ) p (\emptyset/ \{c\} ))$

$\hspace{1.7cm}
+ (-p (\emptyset / \{b,c\} ) p (\emptyset/ \{a\} ) + p (\emptyset / \{c\} ) p (\emptyset / \{a\} ) p (\emptyset/ \{b\} ))$

\

$\hspace{1.3cm} =   p (\emptyset /  \{a,b,c\} ) -p (\emptyset /  \{a\} ) p (\emptyset /  \{b\} ) p (\emptyset /  \{c\} )$

$\hspace{1.7cm} -p (\emptyset/ \{c\} ) (p (\emptyset/ \{a,b\} ) -p (\emptyset/ \{a\} ) p (\emptyset/ \{b\} ))$

$\hspace{1.7cm}
-p (\emptyset / \{b\} ) (p (\emptyset / \{a,c\} ) -p (\emptyset /\{a\} ) p (\emptyset    /  \{c\} ))$

$\hspace{1.7cm}
-p (\emptyset /  \{a\} )( p (\emptyset /  \{b,c\} ) -p (\emptyset /  \{b\} ) p (\emptyset /  \{c\} ))$

\

$\hspace{1.3cm} =  (p (\emptyset/ \{a,b,c\} ) -p (\emptyset/ \{a\} ) p (\emptyset    / \{b\} ) p (\emptyset/ \{c\} ))$

$\hspace{1.7cm}-p (\emptyset    /  \{c\} ) D_{\{a.b\} }$

$\hspace{1.7cm}
-p (\emptyset    /  \{b\} ) D_{\{a,c\}}   $

$\hspace{1.7cm}
-p (\emptyset    /  \{a\} ) D_{\{b,c\} }$

\item  $ D'_\emptyset    = D_\emptyset    = 1$.  We use $D'_A =   \sum_{B<A} R(B/A) D_B   D_{A-B}$.

\item  $D'_{\{a\}}  =  R (\emptyset/  \{a\} ) D_\emptyset  D_{\{a\}} 
+ R (  \{a\}  /  \{a\} ) D_{\{a\}} D_\emptyset $

$\hspace{0.9cm}   = 0$

\item $ D'_{ \{a,b\}}  = \sum_{B<A} R (B /  \{a,b\} ) D_B D_{ \{a,b\} - B} $

$\hspace{1.1cm}  = R (\emptyset/ \{ a,b\} ) D_\emptyset   D_{ \{a,b\}} 
+ R ( \{a\} / \{a,b\} ) D_{\{a\}} D_{\{b\}} $

 $\hspace{1.1cm} 
+ R ( \{b\} /  \{a,b\} ) D_{ \{b\}} D_{\{a\}} 
+ R ( \{a,b\} / \{a,b\} ) D_{\{a,b\}} D_\emptyset  $

 $\hspace{1.1cm} 
= R ( \emptyset /  \{a,b\} ) D_{ \{a.b\}}  + R (  \{a,b\} /  \{a,b\} ) D_{\{a.b\}} D_\emptyset$

$\hspace{1.1cm} 
= (R (\emptyset /  \{a,b\} ) + R (  \{a,b\} /  \{a,b\} )) D_{ \{a,b\}} $

\item   $D'_{\{a.b.c\}} 
= \sum_{B<A} R (B /  \{a,b,c\} ) D_B D_{\{a.b.c\} -B}$

 $\hspace{1.4cm}= R ( \emptyset /  \{a,b,c\} ) D_\emptyset D_{\{a.b.c\}}  + R (  \{a\}  /  \{a,b,c\} ) D_{\{a\}}  D_{\{b,c\} }$

$\hspace{1.6cm}
+ R(  \{b\}  /  \{a,b,c\} ) D_{\{b\}}  D_{\{a,c\}}  + R (  \{c\}  /  \{a,b,c\} ) D_{ \{c\}}  D_{\{a,b\}} $

$\hspace{1.6cm}
+ R( \{a,b\} /  \{a,b,c\} ) D_{\{a.b\}} D_{\{c\}} + R ( \{a,c\} / \{a,b,c\} D_{\{a,c\}} D_{\{b\}}$

$\hspace{1.6cm}
+ R (  \{b,c\}  /  \{a,b,c \} ) D_{\{b,c\}}  D_{\{a\}}  + R (  \{a,b,c\} /  \{a,b,c\} ) D_{\{a.b.c\}}  D_\emptyset $

$\hspace{1.4cm}=  R(\emptyset/\{a,b,c\} ) D_{\{a,b,c\}}  + R (  \{a,b,c\}  /  \{a,b,c\} ) D_{\{a,b,c\}}$

$\hspace{1.4cm}=  
(R(\emptyset /  \{a,b,c\} ) + R (  \{a,b,c\}  /  \{a,b,c\} )) D_{\{a,b,c\}} 
$
\end{enumerate}

Using these examples we have no trouble with the universal set N, for
it can be any set. We have, moreover, shown that for gametes with
zero, one, two or three loci, the value of the gametic disequilibrium
in the offspring generation of a given gamete is equal to the product
of its gametic disequilibrium in the given generation, multiplied by
the probability of no recombination among loci in the gamete.

\section{FIXED POINTS AS LIMIT POINTS}

We have developed  formulas to calculate evolution of transformed frequencies, fixed points of the dynamics  and  gametic disequilibrium. We discovered that each initial condition gives rise to a fixed point of the dynamics and so we reasonably expect  that  each initial condition is absorbed by its corresponding fixed point. Nevertheless, things are not that easy:

\

Let us, Dear Reader, confront you with the next \textit{objection}: fixed points cannot be limit points of the dynamics because the recurrent formula for the evolution of transformed frequencies (26) has the same form as that for the evolution of disequilibrium (45). In fact we have:

$$t' (A) =\sum_{B<A} R_{B/A}t (B)t(A-B)	$$

$$ D'_A =   \sum_{B<A} R(B/A) D_B   D_{A-B} $$

Therefore, were the fixed points be the limit points of the dynamics, then the disequilibrium would tend to zero. But, because the transformed frequencies evolve according to the same law, transformed frequencies would also tend to zero and so the population would disappear. Hence, the objection predicts that fixed points are isolated  from the general evolution of  the diverse frequencies.

\

We expect you to have a lot of fun trying to make concepts clear. Hint: the first step to solve this problem is to explicitly calculate the disequilibrium after an arbitrary number of generations for a number of loci equal to 1, 2 or 3 and to contrast these values with the corresponding ones for the transformed frequencies.

\section{THE EFFECT OF MIGRATION}

Until now we have dealt with a panmictic population. In this
section we would like to introduce demification  into the population
with a general scheme of migration (a deme is a subpopulation). Let us show that the
migration   of individuals originate the same change on relatives frequencies of individuals and of gametes.  

Let us model migration as a discrete operation whose unit of time coincides with that of reproduction. Before migration, the breeding population in deme number $i$ is $n_i$ and let $m_{ij}$ be the migration rate per individual from deme $i$ to $j$. Then, after migration, the population $n^m_i$ is:

$n^m_i = n_i - \sum_{j\ne i} m_{ij}n_i + \sum_{j\ne i} m_{ji}n_j$

where we have included an outflow and an inflow caused by migration. Organizing, we get:

$$
	 n^m_i = (1-   \sum_{j\ne i} m_{ij}) n_i + \sum_{j\ne i} m_{ji} n_j 
$$

Let us consider now the effect of migration over genotype frequencies. Before
migration, the frequency of individuals with genotype $(B,C)$
at deme $i$ is $p_i (BC) = p_i (B)p_i(C)$ and after migration is $p^m_i (B)p^m_i (C)$.
Likewise, the
equation that includes the effects of migration  is: 

\

$p^m_i (B)p^m_i (C) = (1 -\sum_j    m_{ij})
p_i (B)p_i (C) + \sum_j   m_{ij}p_j (B)p_j (C)$.

\

 Summing up over $C<N$,
factoring terms with gamete $B$ and recalling that

$\sum    p^m_i (C) = \sum p_i (C) = \sum p_j (C) =1 $,  
we get

\begin{equation}
	p^m_i(B) = (1 -\sum    m_{ij}) p_i (B) + \sum m_{ji}p_j (B)  
\end{equation}

This equation can be multiplied at both sides by $\S  (A\cap B)$ and then
summed up as $B < N$, then

\

$\sum_{B<N} \S  (A\cap B)p^m_i(B) = (1- \sum_{i\ne j} m_{ij})  \sum_{B<N} \S  (A \cap B)p_i (B)$

$\hspace{4.4cm}
+ \sum_{i\ne j} m_{ji} \sum_{B<N}    \S  (A \cap B) p_j (B)$

\

Since the sum is given over $B< N$ we can recognize here an equation
regulating the migration of transformed frequencies:

\begin{equation}
	t^m_i(A) = (1-\sum m_{ij})t_i(A) + \sum m_{ji} t_j (A)
\end{equation}

We see that migration of individuals, of
gametes, and of transformed frequencies, all have the same form. This property can be generalized to the marginal probabilities too, for they are sums of some normal
frequencies, but equilibrium frequencies require special treatment
given by (35) and (50):

\begin{equation}
	e^m_{i B} = \prod_{b\in B} [ (1 -\sum    m_{ij}) t_{ib} + \sum m_{ij} t_{jb}] 
\end{equation}

A model involving  reproduction and migration can be formulated if we update equation (50), where primes mean the offspring generation:

\

\begin{equation}
	t^{'m}_i(A) = (1-\sum m_{ij})t'_i(A) + \sum m_{ji} t'_j (A)
\end{equation}

\

In order to see the effect of migration,   it is
necessary to transform this equation in an expression containing disequilibrium
terms. To this aim, let us elaborate the  term:

  $ t' (A) = \sum R(B/A)t(B)t(A-B)  $
  
   by replacing $t(B)$ by $d_B + e_B$ and $t(A-B)$ by $d_{A-B} + e_{A-B}$, we have:

$t'(A) = \sum R(B/A) (d_B + e_B)(d_{A-B} + e_{A-B} ) $

$t'(A) = \sum R(B/A) (d_Bd_{A-B} + d_Be_{A-B} +  e_Bd_{A-B} + e_B e_{A-B} ) $

since $e_Be_{A-B} = e_A $ and $ \sum_{B<A}  R(B/A) = 1$, then

\begin{equation}
	 t'(A) = \sum R(B/A) (d_B d_{A-B} + d_Be_{A-B} +  e_Bd_{A-B} ) + e_A
\end{equation}

 In particular, for any deme, the disequilibrium after reproduction
alone is $d'_A = t' (A)-e_A$ because reproduction does not change
equilibrium frequencies. Then

\begin{equation}
	d'_A = t'(A) - e_A = \sum R(B/A) (d_Bd_{A-B} + d_Be_{A-B} +  e_Bd_{A-B}  )
\end{equation}

 The disequilibrium after one round of reproduction plus migration is given by:

$ d^{'m}_{i(A)} =  t^{'m}_i (A) -e^m_{ i(A)}  $ 
	
Invoking   (52),   	we get:

$d^m_{i(A)} = (1-\sum m_{ij})t'_i(A) + \sum m_{ji} t'_j (A) -e^m_{ i(A)}$

Recalling (51) and (53), we have:
 
\begin{equation}
d^m_{i(A)} = (	1 -\sum m_{ij}) (\sum  R(B/A) (d_{i(B)} d_{i(A-B)} + d_{i(B)} e_{i(A-B)} +  e_{i(B)} d_{i(A-B)} +e_{i(A)} )) $$

$$\hspace{1.5cm}   + \sum m_{ij} (\sum  R(B/A) ) (d_{j(B)} d_{j(A-B)}+ d_{j(B)} e_{j(A-B)} +  e_{j(B)} d_{j(A-B)} +e_{j(A)}  )  $$

$$\hspace{1.5cm}  - \prod_{b\in A} [(1-\sum_j m_{ij}) t_{ib} + \sum_j m_{ji} t_{jb} ]$$

$$sums \hspace{0.2cm} over  \hspace{0.2cm} j \ne i,  \hspace{0.2cm} and  \hspace{0.2cm} B<A 
\end{equation}

Hence, when there is no gametic disequilibrium prior to one round of
reproduction + migration $d^m_{i(A)}$ reduces to:

\begin{equation}
d^m_{i(A)} = (	1 -\sum_j m_{ij})  e_{i(A)} + \sum_j m_{ji}  e_{j(A)}  - \prod_{b\in A} [(1-\sum_j m_{ij}) t_{ib} + \sum_j m_{ji} t_{jb} ] 
\end{equation}

To expand the productore, let us note that

\

$(m_1 + n_1 ) (m_2 + n_2) = m_1 m_2 + m_1 n_2 + n_1 m_2 + n_1 n_2$

$(m_1 + n_1)(m_2 +n_2)(m_3 +n_3)=m_1m_2m_3 +m_1n_2m_3 +n_1m_2m_3  
+n_1n_2m_3$

$\hspace{5.3cm}  + m_1m_2n_3 + m_1n_2n_3 + n_1m_2n_3 + n_1n_2n_3$

$\hspace{4.7cm}  = m_1m_2m_3 +m_2m_3n_1 +m_1m_3n_2 + m_1m_2n_3   
$

$\hspace{5.3cm} +m_3n_1n_2+ m_1n_2n_3 + m_2n_1n_3 + n_1n_2n_3$
 
This can be generalized to:

\begin{equation}
	\prod_{b \in N} (m_b +n_b) = \sum_C \prod_{b\in H} m_b \prod_{c\in K}n_c = \sum_{H<K} \prod_{b\in H} m_b \prod_{c\in N-H} n_c
\end{equation}
 
 where $C$ is the condition expressed by $H\cup K = N$ and $H\cap K = \emptyset$.
 
 Making $m_b = (1-\sum_j m_{ij})t_{ib}$ and $n_b = \sum m_{ji} t_{jb}$ we have:
 
 \

 $e^m_{i(A)} = \prod_{b\in A} [(1-\sum_j m_{ij})t_{ib} + \sum_j m_{ji} t_{jb} ] $
 
 $\hspace{0.8cm} = \sum_{H<A} \prod_{b\in H} (1-\sum_j m_{ij})t_{ib} \prod_{c\in A-H} ( \sum_j m_{ji} t_{jc})$
 
  $\hspace{0.8cm} = \sum_{H<A}  (1-\sum_j m_{ij})^{\#H }\prod_{b\in H}t_{ib} \prod_{c\in A-H} ( \sum_j m_{ji} t_{jc})$

\begin{equation}
	 e^m_{i(A)} = \sum_{H<A}  (1-\sum_j m_{ij})^{\#H } e_{i(H)} \prod_{c\in A-H} ( \sum_j m_{ji} t_{jc}) 
\end{equation}

To expand $ \prod_{c\in A-H} ( \sum_j m_{ji} t_{jc}) $ , we can generalize (57) to

\begin{equation}
	 \prod_{c\in A-H} ( \sum_j m_{ji} t_{jc})   = \sum_C \prod   m_{1i} t_{1c}    \prod   m_{2i} t_{2c}.. ..\prod   m_{Li} t_{Lc_L}
\end{equation}

 where $C$ means: $c_1 \in K_1$, $c_2 \in K_2$,.. .. $c_L \in K_L$ with $\cup K_i = A-H$ and $K_i \cap K_j = \emptyset $ if $i\ne j$.
 
 Equation (59) can be elaborated to

	 $\prod_{c\in A-H} ( \sum_j m_{ji} t_{jc})   = \sum_C     m^{\#K_1}_{1i}  m^{\#K_2}_{2i}   .. ..m^{\#K_L}_{Li}      \prod_{c_1\in K_1}   t_{1c_1}.. ..  \prod_{c_l\in K_l}   t_{Lc_L} $

\begin{equation}
	\prod_{c\in A-H} ( \sum_j m_{ji} t_{jc})   = \sum_C     m^{\#K_1}_{1i}  m^{\#K_2}_{2i}   .. ..m^{\#K_L}_{Li}    e_1(K_1)e_2(K_2).. .. e_L(K_L)
\end{equation}

Turning back to (58) and them to (56), we get that the gametic disequilibrium created by migration when there was no gametic disequilibrium prior to one round of reproduction + migration is:

\begin{equation}
	d^m_{i(A)} = (1-\sum_j m_{ij})e_{i(A)} + \sum_j m_{ji} e_{j(A)} $$
	
	$$- \sum_{H<A} (1-\sum_j m_{ij})^{\#H} e_{i(H)}\sum_C m^{\#K_1}_{1i} m^{\#K_2}_{2i}.. ..m^{\#K_L}_{Li}e_{i(K_1)}.. ..e_{L(K_L)}
\end{equation}

Our point in that $d^m_i(A)$ in (61) is in general expected to be different
than zero. Therefore, we conclude
that migration can create gametic disequilibrium from zero.
 
\section{CONCLUSION}

Mating among diploid individuals can be reduced to mating
among haploid gametes in a common reservoir. Evolution of gamete
frequencies under any scheme of recombination can be calculated.
The system has fixed points depending on initial conditions, determined
by the so called marginal probabilities. There exists a measure
of gametic disequilibrium, i.e., a function that relates the actual
frequencies to those of equilibrium, the Bennet measure, whose
recurrence equation has the same form as the recurrence equation of
gamete frequencies when they are written in the system of the so
called transformed frequencies. Migration has
the same form for individuals, gametes, transformed and marginal
frequencies. Migration alone can create disequilibrium from zero.

\section{TO KNOW MORE}

\begin{itemize}
	\item The theory has been extended   (Christiansen, 1999). 
	\item How to  simulate recombination in Java with or without bitsets  (Rodr\'\i guez, 2009).
	\item The mathematical theory of the genetics of populations is a well developed discipline. We have classics (Crow and Kimura, 1970) and modern views (Christiansen, 2008).
 
\end{itemize}

\section{BIBLIOGRAPHY}

$\hspace{0.4cm}$ -CHRISTIANSEN, F.B. (1987). The deviation from linkage equilibrium with
multiple loci varying in a stepping -stone cline. \textit{Journal of Genetics}, 66: 45-67.

-CHRISTIANSEN, F.B. (1999). \textit{Population Genetics of Multiple Loci.}
John Wiley \& Sons, Ltd., Chichester, XIV+365pp.

-CHRISTIANSEN, F.B. (2008) \textit{Theories of Population Variation in Genes and Genomes.} Princeton University Press.

-CROW J.F., KIMURA M.  (1970) \textit{Introduction to Population genetics theory.} Burgess Publishing Company.

-KARLIN, S. \& LIBERMAN, U. (1978). Classifications and comparisons of
multilocus recombination distributions. \textit{Proceedings of the National Academy
of Sciences U. S. A.} ,75: 6332- 6336.

-RODRIGUEZ J, F. B. CHRISTIANSEN, H.F. HOENIGSBERG (1988) Theoretical review  on fundamentals of mendelian gametic recombination and migration with n-loci in amphimictic organisms. \textit{ Evoluci\'on biol\'ogica}, Vol 2, Number 2: 177-219. 

-RODRIGUEZ, J. (2009) \textit{Java for the study of evolution}. Recombination is introduced in  Vol 1 pag 78; Recombinant operators appear in Vol 2 pag 123; BitSets are defined in Vol 3, pg 159.   \\ http://www.evoljava.com (Cited: 2-II-2009).

-SCHNELL, F.W. (1961). Some general formulation of linkage effects in inbreeding.
\textit{Genetics}, 46: 947-957.

\section{Glossary}

$\hspace{0.4cm}$  \textbf{Alleles} Variant forms of genes occurring at the same locus are said to be alleles of one another.

\textbf{Amphimictic } That use both sexes in reproduction.

\textbf{Bit} Unit of information that corresponds to a yes else no answer. A bit is usually encoded by 1 else 0.

\textbf{Bitset} A set that represents a binary number.

\textbf{Chromosome} Tiny rods in the cell that carry the genetic information.

\textbf{Deme} A subpopulation that enjoys more or less identity and independence.

\textbf{Diploid} An organism whose cells contain  chromosomes by pairs, one from the mother and one from the father.

\textbf{Disequilibrium measure} A function that calculates the difference between actual state and that of equilibrium.

\textbf{Gamete} Sexual cell able to unite with other in reproduction. Sexual cells have only one version or allele of the genetic information. Here, a gamete is represented by a binary number to  inform, say,  whether in each locus the information comes from the mother else from the father. 

\textbf{Gene} The inheritable information that encodes for certain property. A portion of DNA that encodes for a protein or an enzyme or a part of it (an enzyme is a protein  molecule that selectively accelerates a   reaction in the cell).

\textbf{Haploid} A cell that contains only one copy of genes, as gametes.

\textbf{Homozygote} The quality of a diploid individual of having in the two chromosomes the same information for a given locus.

\textbf{Heterozygote} The quality of a diploid individual of having in the two chromosomes two different versions for a given locus.

\textbf{Linkage} The quality of being together. In nature, when two genes are close one to another, recombination that separates them is less probably.

\textbf{Locus} Here, the specific position occupied by a bit in a binary number that represents a gamete. In biology: the specific position occupied by a gene in a chromosome.

\textbf{Loci} The plural form of loci.

\textbf{Meiosis} Cellular division that at the end gives rise to gametes. 

\textbf{Mendel's law} Working with pea plants in the garden of his monastery, Gregor Mendel made the first model of genetics: First law: Reproduction results from fusion of gametes and gametes can carry only one type of inheritable information  while organisms may carry two. Second law:  for two characteristics the inheritable factors are inherited independently. The first law is correct for diploid organisms, the second is correct only when the factors are not linked because of physical closeness. In the present work, both laws of Mendel are assumed to be true. 

\textbf{Panmictic population} One that has no reproductive barriers or biases with respect to random mating.

\textbf{Parthenogenesis} Optional reproduction of females without the cooperation of males.

\textbf{Recombinant operators} A function that to each pair of gametes associates a third one as a result of recombination.

\textbf{Recombination} A process in which a new combination  of alleles is formed beginning from two previous ones.

\textbf{Zygote} The cell that results from the fusion of the ovule and spermatozoon.  
 
\end{document}